\documentclass[a4paper, aps, prc, showpacs, superscriptaddress, preprintnumbers, nofootinbib, twocolumn]{revtex4-2}
\usepackage[pdftex]{graphicx, color}
\usepackage{graphbox, amsmath, amssymb, amscd, latexsym, bm, braket}
\usepackage[colorlinks=true,pdfstartview=FitV,linkcolor=blue,citecolor=magenta,urlcolor=blue,bookmarks=true,bookmarksnumbered=true]{hyperref}
\newcommand{\sqsNN}{\sqrt{s_{_{\;\!\!N\;\!\!\;\!\!N}}}}

\begin{document}
\title{Spacetime profile of electromagnetic fields in intermediate-energy heavy-ion collisions}

\author{Hidetoshi Taya}
\email{h\_taya@keio.jp}
\address{Department of Physics, Keio University, 4-1-1 Hiyoshi, Kanagawa 223-8521, Japan}
\address{iTHEMS, TRIP Headquarters, RIKEN, Wako 351-0198, Japan}

\begin{abstract}
I numerically estimate the spacetime profile of the electromagnetic fields produced in intermediate-energy heavy-ion collisions at $\sqsNN=2.4\,\mathchar`-\,7.7\;{\rm GeV}$ for a wide range of impact parameters $b=0\,\mathchar`-\,9\;{\rm fm}$, using a hadronic cascade model, JAM (Jet AA Microscopic transport model).  I demonstrate that (1) the produced electromagnetic fields are as strong as $eE, eB = {\mathcal O}((50\;{\rm MeV})^2)$; (2) that the produced fields extend in spacetime about $V_4 = {\mathcal O}((10\;{\rm fm})^4)$; (3) that the magnetic field dominates around the collision point, while the other broad regions of spacetime are dominated by the electric field; and (4) that a topological electromagnetic-field configuration such that ${\bm E}\cdot {\bm B} \neq 0$ is realized.  
\end{abstract}

\maketitle

\section{Introduction}

The possibility of generating strong electromagnetic fields in heavy-ion collisions has been discussed intensively over the past decade for RHIC (Relativistic Heavy Ion Collider) and LHC (Large Hadron Collider) energy scales, where the collision energy $\sqsNN$ (i.e., the center-of-mass energy per nucleon pair) is $\sqsNN = {\mathcal O}(100\,\mathchar`-\,5000\;{\rm GeV})$~\cite{Hattori:2016emy}.  Various transport-model simulations have been carried out, and it has been established that a very strong magnetic field is created in non-central collisions due to the Amp\`{e}re law~\cite{Voronyuk:2011jd, Bzdak:2011yy, Deng:2012pc}.  The magnitude of the magnetic field can reach $eB = {\mathcal O}((0.1\,\mathchar`-\,1\;{\rm GeV})^2)$, which is the strongest field known in the present Universe, but it is at the same time extremely short-lived with lifetime of the order of ${\mathcal O}(0.01\,\mathchar`-\,0.1\;{\rm fm}/c)$ (although it has been argued that finite electric conductivity could make the lifetime longer~\cite{Tuchin:2010vs, McLerran:2013hla, Gursoy:2014aka, Tuchin:2015oka, Li:2016tel, Stewart:2021mjz, Benoit:2025amn}).  The generation of such a strong magnetic field has attracted a great deal of attention, in connection to, e.g., the chiral magnetic effect~\cite{Kharzeev:2007jp, Fukushima:2008xe}, the phase diagram of quantum chromodynamics (QCD) in a magnetic field~\cite{Andersen:2014xxa, DElia:2021yvk, Adhikari:2024bfa}, and non-linear quantum electrodynamics (QED)~\cite{DiPiazza:2011tq, Fedotov:2022ely}.  The strong magnetic field can also leave intriguing experimental signatures.  Such signatures have been observed in actual experiments recently, e.g., the first observations of light-by-light scattering~\cite{ATLAS:2017fur}, the linear Breit-Wheeler process~\cite{STAR:2019wlg, Brandenburg:2022tna}, and charged directed flow~\cite{Gursoy:2014aka, Gursoy:2018yai, STAR:2023jdd, Dash:2024qcc}.  

The past study of the strong electromagnetic field in heavy-ion collisions was motivated by high-energy experiments, i.e., those at RHIC and the LHC.  It is, however, one of the research trends of heavy-ion physics to lower the energy scale to investigate the {\it intermediate} energy regime such that $\sqsNN = {\mathcal O}(2\,\mathchar`-\,10\;{\rm GeV})$.  Compared to the high-energy regime, which is suitable for creating hot baryon-less matter, the intermediate-energy regime is advantageous for creating dense baryonic matter~\cite{Taya:2024zpv}.  To explore such a dense baryonic matter, or dense QCD physics broadly speaking, various experimental programs of intermediate-energy heavy-ion collisions are now being conducted [e.g., the Beam-Energy Scan (BES) program at RHIC~\cite{Aparin:2023fml} and NA61/SHINE at the Super Proton Synchrotron (SPS)~\cite{NA61}] and will be done worldwide (e.g., NICA~\cite{NICA}, FAIR~\cite{FAIR}, HIAF~\cite{HIAF}, and J-PARC-HI~\cite{J-Parc-HI})~\cite{Galatyuk:2019lcf}.  

Given the increasing interest in intermediate-energy heavy-ion collisions, it is natural to ask what the electromagnetic-field profile looks like in this energy regime.  Compared to high-energy collisions, relatively few studies have explored this topic~\cite{Ou:2011fm, Sun:2019hao, Wei:2021yiy, Taya:2024wrm, Panda:2024ccj, Siddique:2025tzd}.  Indeed, physical processes are much more involved due to the baryon stopping, while the system is almost transparent in the high-energy limit, meaning that intermediate-energy heavy-ion collisions require a more sophisticated microscopic treatment of the dynamics.  

Understanding the electromagnetic-field profile is interesting not only because it has potential to induce novel QED/QCD/hadronic processes but also because it can contaminate signals of dense QCD physics.  In fact, electromagnetic observables such as dileptons and photons can in principle be clean probes of dense QCD physics, since they are transparent to the strong interaction.  It is, thus, natural to look for such electromagnetic signatures of dense QCD physics such as the color superconductor phase transition or the conjectured QCD critical point~\cite{Nishimura:2022mku, Savchuk:2022aev, Nishimura:2023oqn, Nishimura:2023not}.  However, these electromagnetic observables are expected to be modified by the presence of a strong electromagnetic field and hence would blur the signals (cf. the impact of the electromagnetic field for the extraction of the symmetry energy~\cite{Wei:2021yiy}).  

In addition to the motivation in heavy-ion physics, understanding of the strong electromagnetic field in intermediate-energy heavy-ion collisions is also of interest to the area of strong-field QED.  In strong-field QED, there is an experimental difficulty that a strong electromagnetic field beyond the critical field strength $eE_{\rm cr} := m_e \approx (0.5\;{\rm MeV})^2$, where $e>0$ is the elementary charge and $m_e$ is the electron mass, cannot be achieved in laboratory experiments.  The strongest electromagnetic field at the present is created with an intense laser, whose focused intensity is $I = 1 \times 10^{23}\;{\rm W/cm^2}$, corresponding to $E \approx 10^{-3}\,E_{\rm cr} $~\cite{Yoon:21}.  Laser intensity is growing rapidly and continuously.  Hence, the current intensity is envisaged to be surpassed by the latest and future facilities such as Extreme Light Infrastructure (ELI), with $I = {\mathcal O}(10^{25}\;{\rm W/cm^2})$~\cite{ELI}, which is, however, still far below the critical field strength $E_{\rm cr}$.  Therefore, a novel method to create a strong electromagnetic field, other than with an intense laser, is highly desirable.  Intermediate-energy heavy-ion collisions can be a natural candidate for this.  

The purpose of this paper is to investigate the spacetime profile of the electromagnetic field produced in intermediate-energy heavy-ion collisions with finite impact parameter.  This work extends the author's previous study~\cite{Taya:2024wrm}, which focused on central collisions and the resulting electric field.  Introducing a finite impact parameter leads to an interplay between electric and magnetic fields.  Namely, in central collisions, the produced field is dominated by the Coulomb field emanating from the collision remnants and is therefore purely electric.  As the impact parameter increases, a circular electric current develops on the reaction plane, generating a magnetic field perpendicular to that plane via Amp\`{e}re’s law.  Consequently, in non-central collisions, both electric and magnetic fields coexist, and their evolution depends sensitively on the collision dynamics, investigation of which requires a realistic transport-model simulation.  

This paper is organized as follow: In Sec.~\ref{sec:2}, I describe the numerical strategy adopted in this work.  Section~\ref{sec:3} presents the numerical results for the estimated electromagnetic field and discusses their physical implications.  A summary and discussion are given in Sec.~\ref{sec:4}.  Additional plots of the spacetime profile of the electromagnetic field are provided in Appendix~\ref{sec:app1}.

{\it Notation}.--- The beam axis and the impact parameter are taken along the $z$- and $x$-directions, respectively.  I use the natural units such that $\hbar = c = 1$.

\section{Numerical method} \label{sec:2}

I estimate the spacetime profile of the electromagnetic field in intermediate-energy heavy-ion collisions with finite impact parameter by using the numerical method developed in Ref.~\cite{Taya:2024wrm}.  In this section, I briefly overview this method.  

I use an event generator, JAM~\cite{Nara:1999dz, JAM}.  JAM is a hadronic cascade model that incorporates resonances, string excitation, and mini-jets as inelastic particle-production mechanisms, allowing it to cover a broad range of collision energies from a few GeV to more than 100\;GeV (see Ref.~\cite{TMEP:2022xjg} for more on various model comparisons).  I ran JAM with the default setting (i.e., no mean field or hydro phase, which can be included as advanced options) in order to provide a baseline result before incorporating such non-trivial physical effects.  

I calculate the electric current $J^\mu$ at each spacetime point $(t, {\bm x})$ from the phase-space distributions of charged particles produced in the collisions obtained by JAM.  For a single collision event, $J^\mu$ is given by
\begin{align}
	J(t,{\bm x}) = \sum_{i \in {\rm all\ charged\ hadrons}} \frac{p^\mu_i}{p^0_i} q_i f(t,{\bm x}) \;, \label{eq:1}
\end{align}
where $p^\mu_i$ and $q_i$ are the four-momentum and the electric charge of the $i$-th hadron at time $t$, respectively, and $f$ is a smearing function to convert the charge $q_i$ to the charge density $q_i f$.  I adopt the relativistic Gaussian smearing function~\cite{Fuchs:1995fa, Oliinychenko:2015lva}, 
\begin{align}
	f(t,{\bm x}) 
	:= \frac{\gamma_i}{(\sqrt{2\pi}\sigma)^3} {\rm e}^{ -\frac{ |{\bm x}-{\bm x}_i|^2 + \gamma_i^2 \left( {\bm v}_i\cdot ({\bm x}-{\bm x}_i) \right)^2}{2\sigma^2} } \;,
\end{align}
where ${\bm x}_i, {\bm v}_i$, and $\gamma_i := 1/\sqrt{1-({\bm v}_i)^2}$ are the position, velocity, and the gamma factor of the $i$-th hadron, respectively, and $\sigma$ is the smearing width, which I fix as 1\;fm.  

Once the electric current (\ref{eq:1}) is obtained, can I calculate the electromagnetic field from the retarded potential, 
\begin{align}
	A^\mu (t,{\bm x}) 
	:= \frac{1}{4\pi} \int {\rm d}^3{\bm x}' \frac{J^\mu(t-|{\bm x}-{\bm x}'|,{\bm x}')}{|{\bm x}-{\bm x}'|} \;, \label{eq:3} 
\end{align}
as
\begin{align}
	{\bm E}_n := -\partial_t {\bm A} - {\bm \nabla} A^0 
	\ \ {\rm and} \ \ 
	{\bm B}_n := {\bm \nabla} \times {\bm A} \;,  \label{eq:4}
\end{align}
where I have added the subscript $n$ to make sure that it is a single-event result at the $n$-th event.  I carry out the integration in Eq.~(\ref{eq:3}) numerically via the Gauss-Legendre quadrature method and the differentiation in Eq.~(\ref{eq:4}) via the center differentiation, and confirmed that both are converging well with the numerical parameters used in the simulation.

In this paper, I am interested in an event-averaged value, rather than a single-event value obtained in Eq.~(\ref{eq:4}).  Therefore, I take the event averaging of the electromagnetic fields as
\begin{align}
	{\bm E} := \frac{1}{N} \sum_{n=1}^N {\bm E}_n
	\ \ {\rm and}\ \ 
	{\bm B} := \frac{1}{N} \sum_{n=1}^N {\bm B}_n \;, \label{eq:5}
\end{align}
where $N$ is the total number of events.  Due to computational constraints, I chose a relatively small event number $N=100$.  While this number is not large enough to completely eliminate event-by-event fluctuations, I have verified that it is sufficient for achieving a well-smoothed electromagnetic-field distribution in spacetime (see Sec.~\ref{sec:3}), and for that the sensitivity to the choice of the smearing width $\sigma$ becomes negligible after the event averaging.

\section{Numerical results} \label{sec:3}

I present the numerical results obtained using the method described in Sec.~\ref{sec:2}, and analyze the electromagnetic fields generated in non-central collisions of gold ions at intermediate energies.  The main findings in this section are as follows: (1) the generated electromagnetic fields reach strengths of $eE, eB = {\mathcal O}((50\;{\rm MeV})^2)$; (2) the fields extend over a spacetime volume of $V_4 = {\mathcal O}((10\;{\rm fm})^4)$; (3) the magnetic field is localized near the collision point, whereas broader regions of spacetime are dominated by the electric field; and (4) a topologically nontrivial electromagnetic configuration with ${\bm E}\cdot {\bm B} \neq 0$ is realized.

\subsection{Spacetime profile}

\begin{figure*}[!t]
\flushleft{\hspace*{15mm}\mbox{$t=1.0\;{\rm fm}/c$ \hspace{15.6mm} $4.5\;{\rm fm}/c$ \hspace{17.6mm} $8.0\;{\rm fm}/c$ \hspace{17mm} $11.5\;{\rm fm}/c$ \hspace{17mm} $15.0\;{\rm fm}/c$}} \\
\vspace*{1mm}
\hspace*{-33mm}
\includegraphics[align=t, height=0.2349\textwidth, clip, trim = 35 0 115 32]{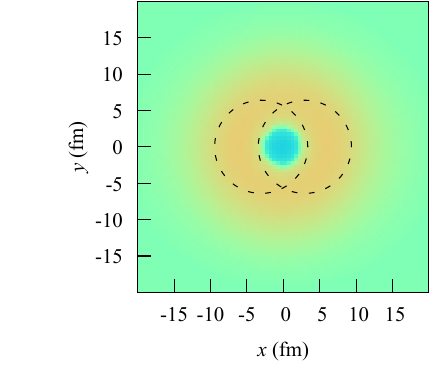}\hspace*{3.3mm}
\includegraphics[align=t, height=0.2349\textwidth, clip, trim = 95 0 115 32]{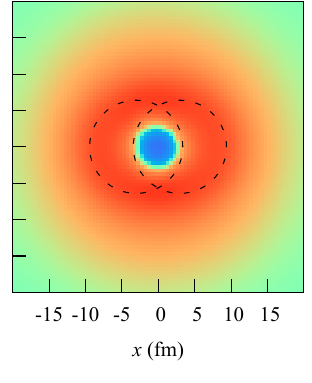}\hspace*{-3.4mm}
\includegraphics[align=t, height=0.2349\textwidth, clip, trim = 95 0 115 32]{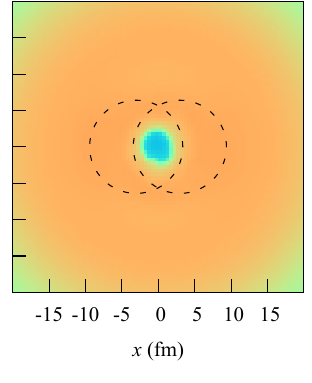}\hspace*{-3.4mm}
\includegraphics[align=t, height=0.2349\textwidth, clip, trim = 95 0 115 32]{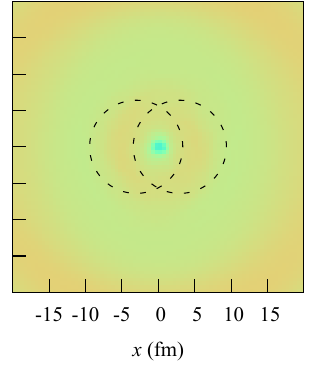}\hspace*{-10.3mm}
\includegraphics[align=t, height=0.2349\textwidth, clip, trim = 95 0  55 32]{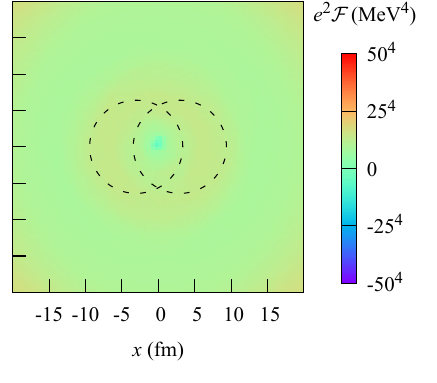}\hspace*{-6mm} \\
\vspace*{2mm}
\hspace*{-33mm}
\includegraphics[align=t, height=0.2349\textwidth, clip, trim = 35 0 115 32]{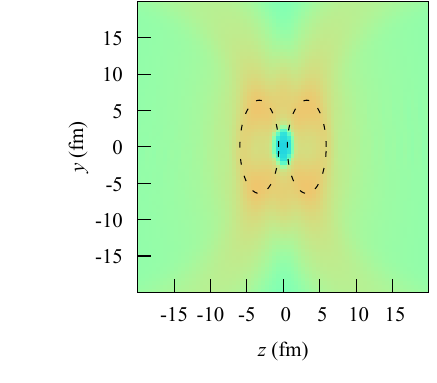}\hspace*{3.3mm}
\includegraphics[align=t, height=0.2349\textwidth, clip, trim = 95 0 115 32]{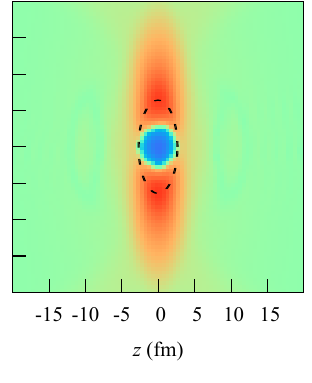}\hspace*{-3.4mm}
\includegraphics[align=t, height=0.2349\textwidth, clip, trim = 95 0 115 32]{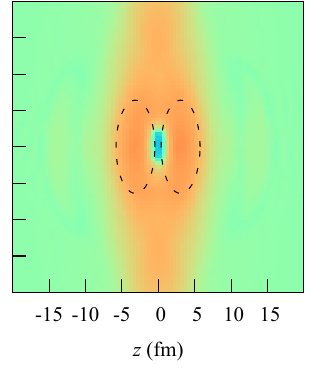}\hspace*{-3.4mm}
\includegraphics[align=t, height=0.2349\textwidth, clip, trim = 95 0 115 32]{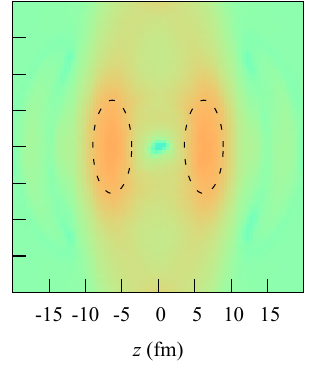}\hspace*{-3.4mm}
\includegraphics[align=t, height=0.2349\textwidth, clip, trim = 95 0 115 32]{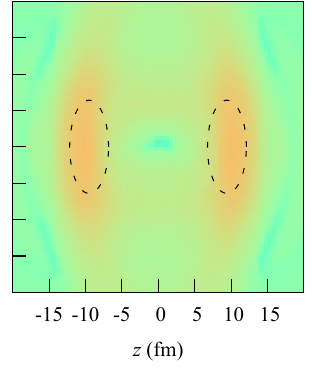}\hspace*{-6mm} \\
\vspace*{2mm}
\hspace*{-33mm}
\includegraphics[align=t, height=0.2349\textwidth, clip, trim = 35 0 115 32]{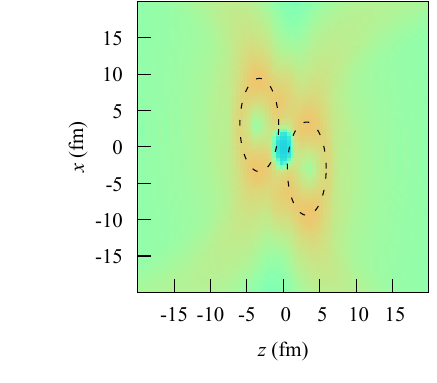}\hspace*{3.3mm}
\includegraphics[align=t, height=0.2349\textwidth, clip, trim = 95 0 115 32]{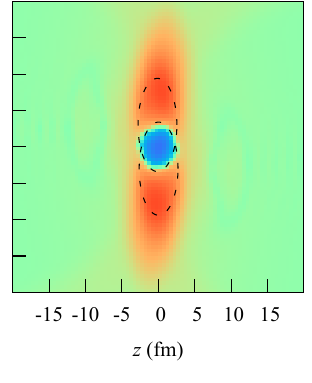}\hspace*{-3.4mm}
\includegraphics[align=t, height=0.2349\textwidth, clip, trim = 95 0 115 32]{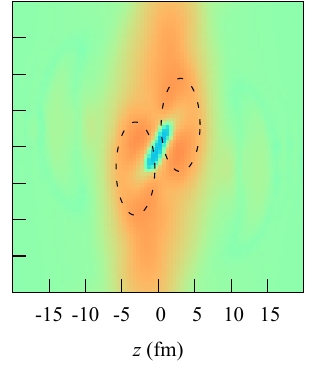}\hspace*{-3.4mm}
\includegraphics[align=t, height=0.2349\textwidth, clip, trim = 95 0 115 32]{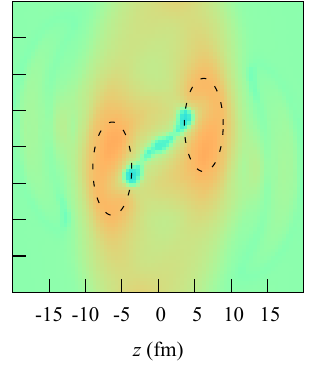}\hspace*{-3.4mm}
\includegraphics[align=t, height=0.2349\textwidth, clip, trim = 95 0 115 32]{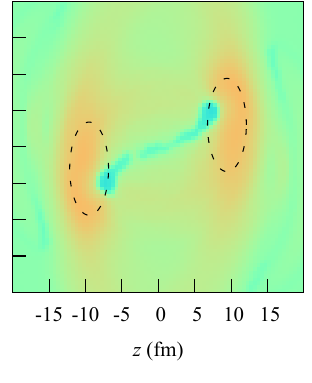}\hspace*{-6mm} \\
\caption{\label{fig:1} The spacetime profile of the electromagnetic Lorentz invariant ${\mathcal F} = {\bm E}^2 - {\bm B}^2$ at a fixed collision energy $\sqsNN=4.5\;{\rm GeV}$ and an impact parameter $b=6\;{\rm fm}$.  The black dashed circles indicate the locations of the colliding ions in the free-streaming limit (i.e., the ions continue their uniform linear motion with the initial beam velocity, without any interactions).  }
\end{figure*}

To build a basic understanding of the electromagnetic field generated in intermediate-energy non-central heavy-ion collisions, I begin by qualitatively examining its spacetime profile through a series of density plots.  For illustrative purposes, the collision energy and impact parameter are fixed here at $\sqsNN = 4.5\;{\rm GeV}$ and $b=6\;{\rm fm}$, respectively; however, the qualitative features remain consistent across other collision energies and impact parameters.  

Figure~\ref{fig:1} shows the spacetime evolution of an electromagnetic Lorentz invariant, 
\begin{align}
	{\mathcal F} := {\bm E}^2 - {\bm B}^2 \;.
\end{align}
See Appendix~\ref{sec:app1} for the same plots for each electromagnetic component $E_x, E_y, E_z, B_x, B_y,$ and $B_z$.  

Before examining the figure, I note that the sign of the invariant ${\mathcal F}$ is important.  It indicates whether the electric or magnetic component of the electromagnetic field dominates.  Namely, the field is {\it electric} for ${\mathcal F}>0$ and {\it magnetic} for ${\mathcal F}<0$.  In the figure, ${\mathcal F}>0$ and ${\mathcal F}<0$ are represented by red and blue colors, respectively.  

Observing Fig.~\ref{fig:1}, one understands that the red region occupies the majority of the spacetime, while the blue region is confined to a small area near the center.  This indicates that the produced electromagnetic fields are primarily dominated by the electric component throughout most of the spacetime, with the magnetic component becoming significant only around the overlapped region of the colliding two nuclei.  The majority of the electric field is simply the consequence of the long-range nature of the Coulomb law.  

One also observes that the produced electromagnetic field is highly anisotropic, i.e., the spatial distribution is strongly squeezed in the beam $z$-direction.  This is a direct consequence of the relativistic Lorentz contraction.  It is notable that such a relativistic effect cannot be dismissed even at the intermediate-energy regime.  

\begin{figure}[!t]
\hspace*{-70mm}
\includegraphics[align=t, width=0.4\textwidth, clip, trim=100 100 100 0]{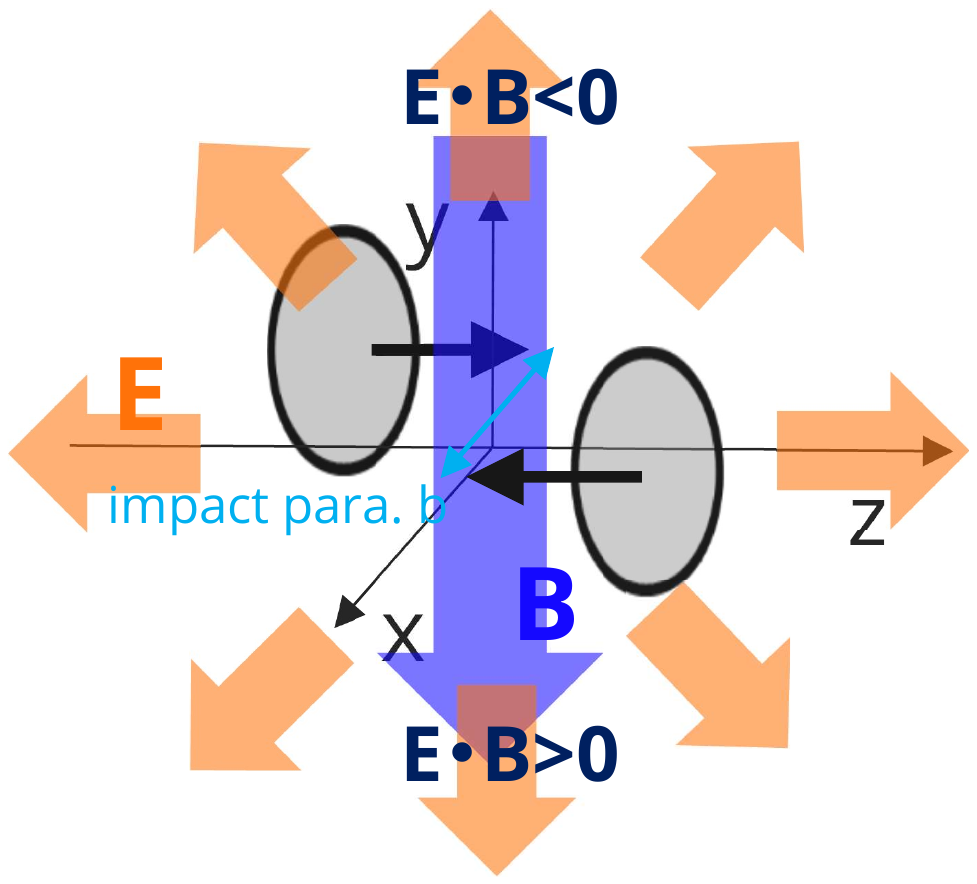}\caption{\label{fig:2} A schematic picture of how nonzero ${\mathcal G}$ emerges.  }
\end{figure}

Next, I consider another electromagnetic Lorentz invariant, 
\begin{align}
	{\mathcal G} := {\bm E} \cdot {\bm B} \;.  
\end{align}
It is an interesting feature that ${\mathcal G}$ can be non-vanishing at intermediate-energy non-central heavy-ion collisions.  An intuitive explanation of how ${\mathcal G}$ becomes nonzero is provided in Fig.~\ref{fig:2}.  There are two physical processes contributing to the nonzero ${\mathcal G}$.  First, the Coulomb electric field is generated by the charged matter formed by the collision participants.  The resulting electric field emanates (roughly) radially from the collision point.  Second, the magnetic field is produced according to Amp\`{e}re's law.  The charged matter is not static but ``rotates" due to the net angular momentum supplied by the finite impact parameter~\cite{Deng:2020ygd}.  As a result, a circulating electric current exists around the $y$-axis, producing a magnetic field along the $-y$ direction.  The resulting field directions are shown in Fig.~\ref{fig:2}.  It is evident that the electric and magnetic fields are not orthogonal to each other, and hence ${\mathcal G} \neq 0$ is concluded.  The figure also indicates that (i) the magnitude of ${\mathcal G}$ becomes the largest (vanishes) on the $y$-axis (at $y=0$), where the Coulomb electric field aligns entirely with the $y$-axis (the $y$-component of the Coulomb field vanishes), and that (ii) the sign of ${\mathcal G}$ flips with that of $y$.

\begin{figure*}[!t]
\flushleft{\hspace*{15mm}\mbox{$t=1.0\;{\rm fm}/c$ \hspace{15.6mm} $4.5\;{\rm fm}/c$ \hspace{17.6mm} $8.0\;{\rm fm}/c$ \hspace{17mm} $11.5\;{\rm fm}/c$ \hspace{17mm} $15.0\;{\rm fm}/c$}} \\
\vspace*{1mm}
\hspace*{-33mm}
\includegraphics[align=t, height=0.2349\textwidth, clip, trim = 35 0 115 32]{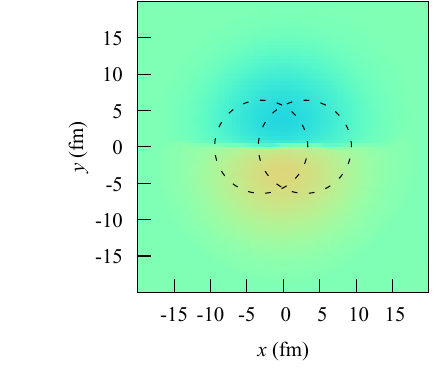}\hspace*{3.3mm}
\includegraphics[align=t, height=0.2349\textwidth, clip, trim = 95 0 115 32]{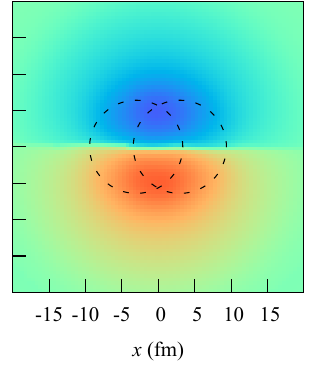}\hspace*{-3.4mm}
\includegraphics[align=t, height=0.2349\textwidth, clip, trim = 95 0 115 32]{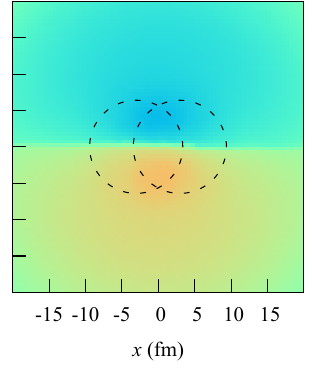}\hspace*{-3.4mm}
\includegraphics[align=t, height=0.2349\textwidth, clip, trim = 95 0 115 32]{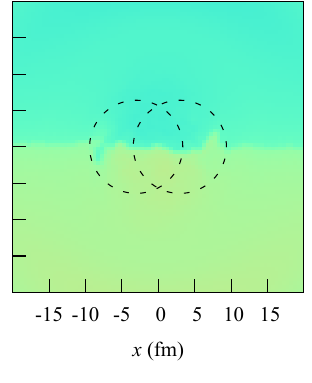}\hspace*{-10.3mm}
\includegraphics[align=t, height=0.2349\textwidth, clip, trim = 95 0  55 32]{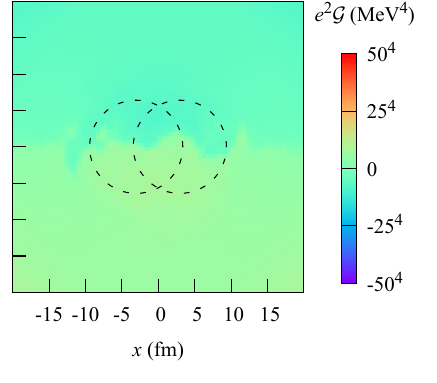}\hspace*{-6mm} \\
\vspace*{2mm}
\hspace*{-33mm}
\includegraphics[align=t, height=0.2349\textwidth, clip, trim = 35 0 115 32]{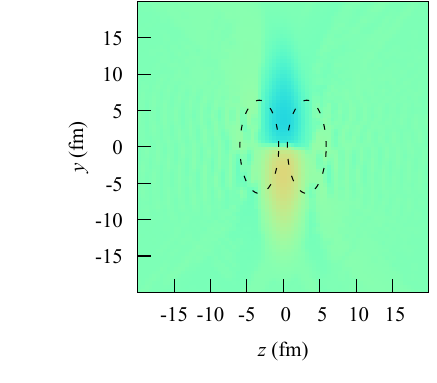}\hspace*{3.3mm}
\includegraphics[align=t, height=0.2349\textwidth, clip, trim = 95 0 115 32]{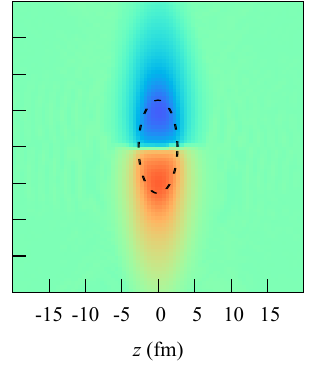}\hspace*{-3.4mm}
\includegraphics[align=t, height=0.2349\textwidth, clip, trim = 95 0 115 32]{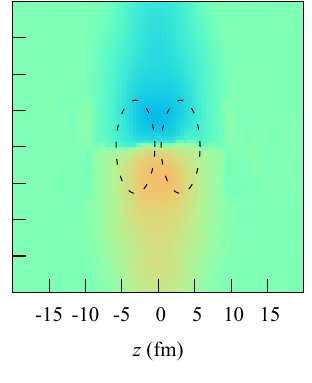}\hspace*{-3.4mm}
\includegraphics[align=t, height=0.2349\textwidth, clip, trim = 95 0 115 32]{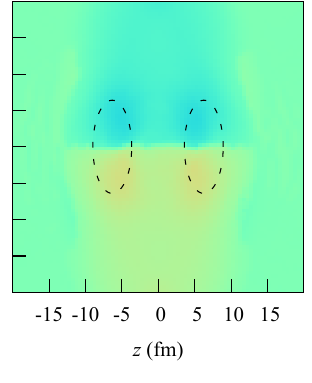}\hspace*{-3.4mm}
\includegraphics[align=t, height=0.2349\textwidth, clip, trim = 95 0 115 32]{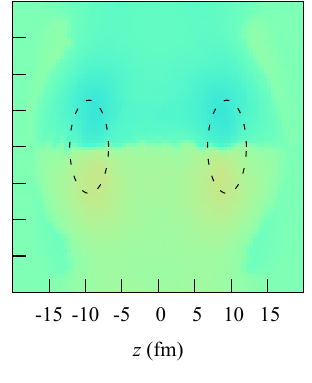}\hspace*{-6mm} \\
\vspace*{2mm}
\hspace*{-33mm}
\includegraphics[align=t, height=0.2349\textwidth, clip, trim = 35 0 115 32]{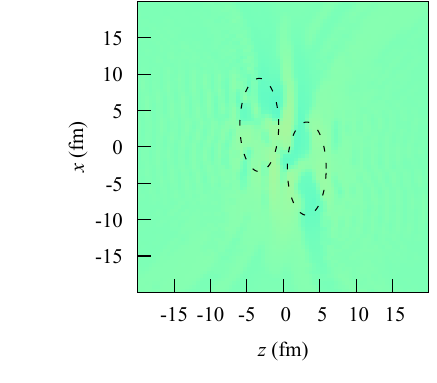}\hspace*{3.3mm}
\includegraphics[align=t, height=0.2349\textwidth, clip, trim = 95 0 115 32]{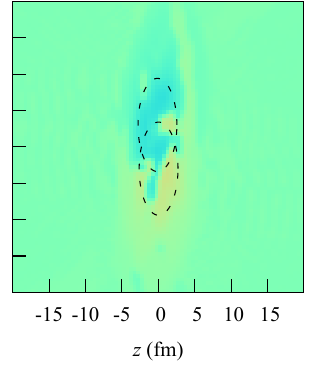}\hspace*{-3.4mm}
\includegraphics[align=t, height=0.2349\textwidth, clip, trim = 95 0 115 32]{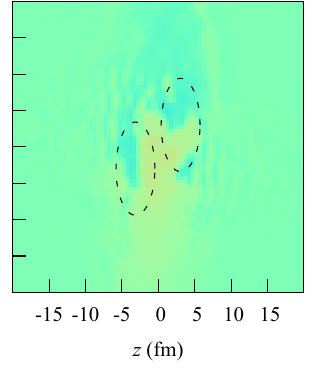}\hspace*{-3.4mm}
\includegraphics[align=t, height=0.2349\textwidth, clip, trim = 95 0 115 32]{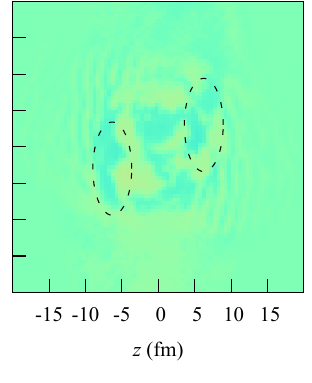}\hspace*{-3.4mm}
\includegraphics[align=t, height=0.2349\textwidth, clip, trim = 95 0 115 32]{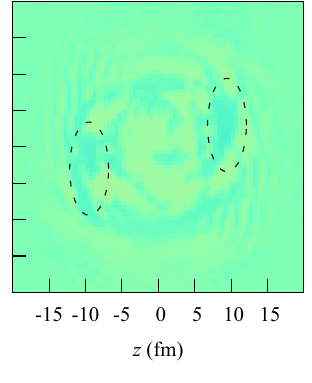}\hspace*{-6mm} \\
\caption{\label{fig:3} The same plot as Fig.~\ref{fig:1} but for the other electromagnetic Lorentz invariant ${\mathcal G} = {\bm E}\cdot{\bm B}$.  }
\end{figure*}

Figure~\ref{fig:3} shows the spacetime evolution of ${\mathcal G}$.  One observes the generation of nonzero ${\mathcal G}$ and that the sign of ${\mathcal G}$ is consistent with the expectation given in Fig.~\ref{fig:2}.  Note that there exists tiny non-vanishing ${\mathcal G}$ at $y=0$ (the bottom of Fig.~\ref{fig:3}), which comes from residual event-by-event fluctuations, and I expect that they will vanish as the event number $N$ increases.

Comparing Fig.~\ref{fig:3} with Fig.~\ref{fig:1}, one notices that the order of the magnitudes of the strength and area of ${\mathcal G}$ do not change significantly from those of ${\mathcal F}$.  This implies that, for a realistic description of the electromagnetic-field physics in intermediate-energy heavy-ion collisions, it is necessary to take into account not only ${\mathcal F}$ but also ${\mathcal G}$, and simple electromagnetic-field models like a purely magnetic-field background or constant crossed-field are insufficient.

\subsection{Time evolution}

\begin{figure*}[!t]
\flushleft{\hspace*{21mm}\mbox{$\sqsNN=3.0\;{\rm GeV}$ \hspace{20mm} $4.5\;{\rm GeV}$ \hspace{27mm} $6.2\;{\rm GeV}$ \hspace{27mm} $7.2\;{\rm GeV}$}} \\
\hspace*{-36.5mm}
\includegraphics[align=t, height=0.189277645\textwidth, clip, trim = 0   45 15 35]{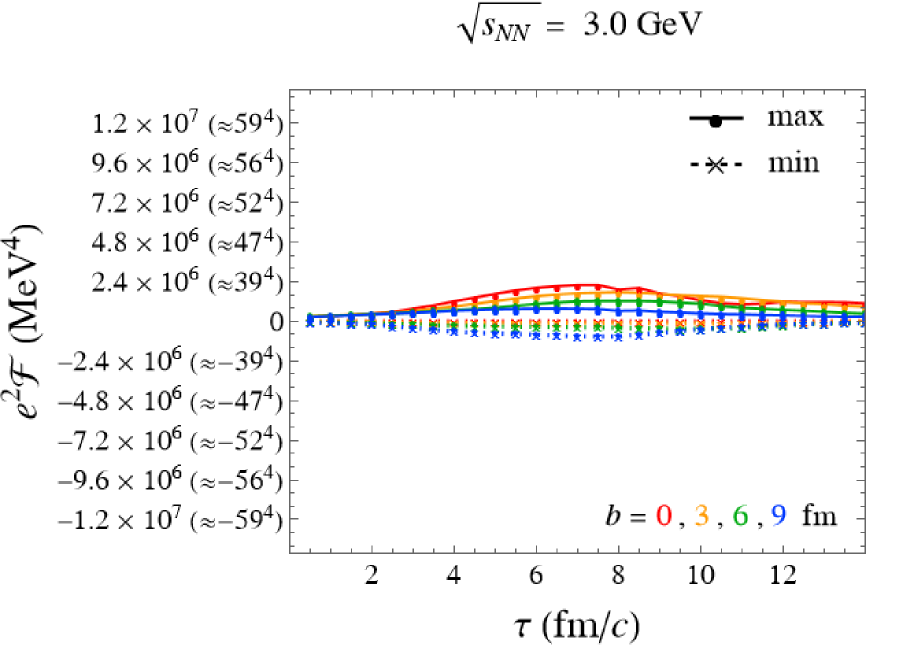}\hspace*{7.8mm}
\includegraphics[align=t, height=0.189277645\textwidth, clip, trim = 139 45 15 35]{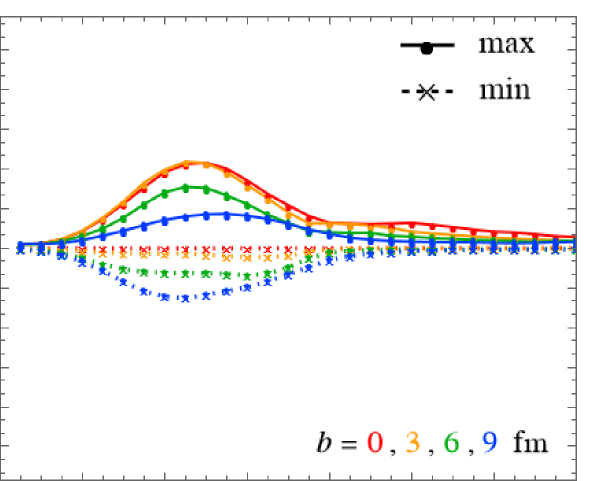}\hspace*{-2.35mm}
\includegraphics[align=t, height=0.189277645\textwidth, clip, trim = 139 45 15 35]{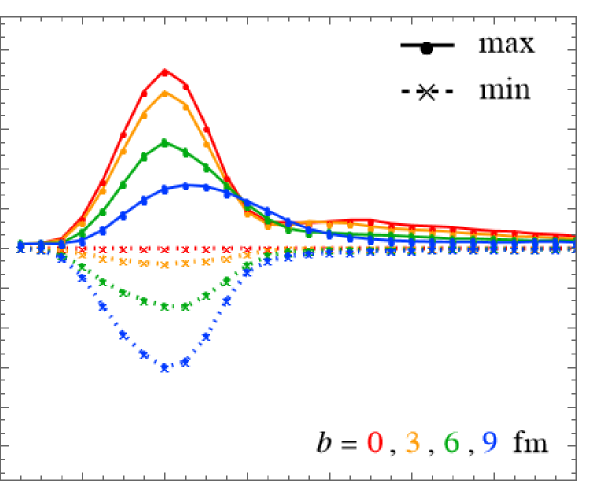}\hspace*{-2.35mm}
\includegraphics[align=t, height=0.189277645\textwidth, clip, trim = 139 45 15 35]{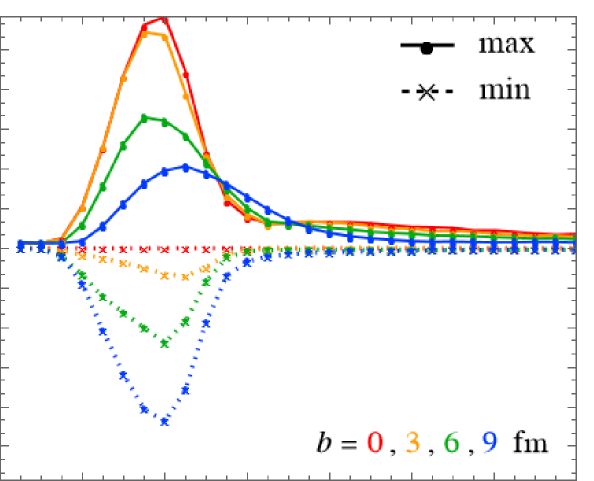} \\
\vspace*{-0.4mm}
\hspace*{-36.5mm}
\includegraphics[align=t, height=0.22\textwidth, clip, trim = 0 0 15 42.5]{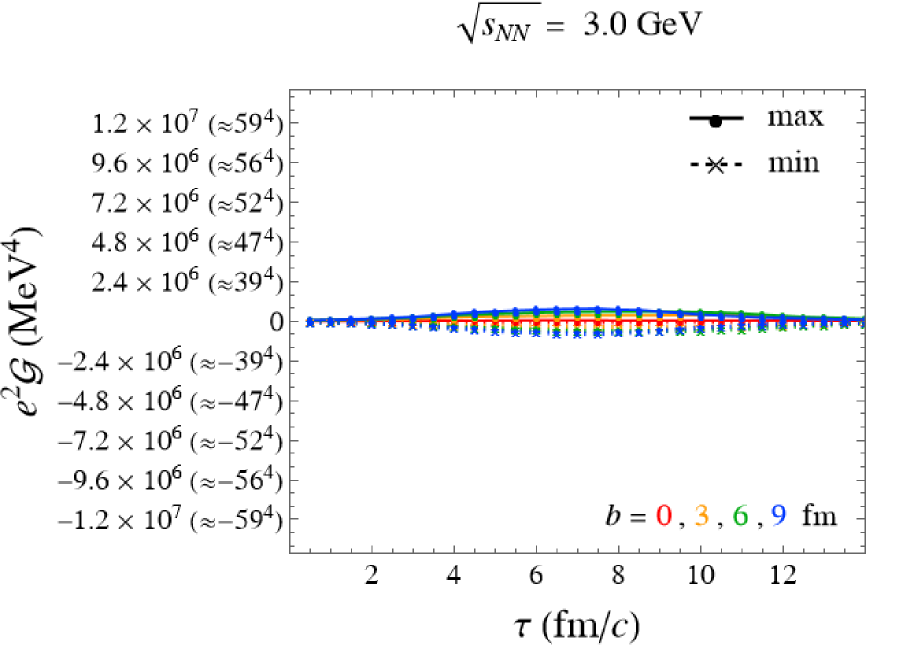}\hspace*{7.8mm}
\includegraphics[align=t, height=0.22\textwidth, clip, trim = 139 0 15 42.5]{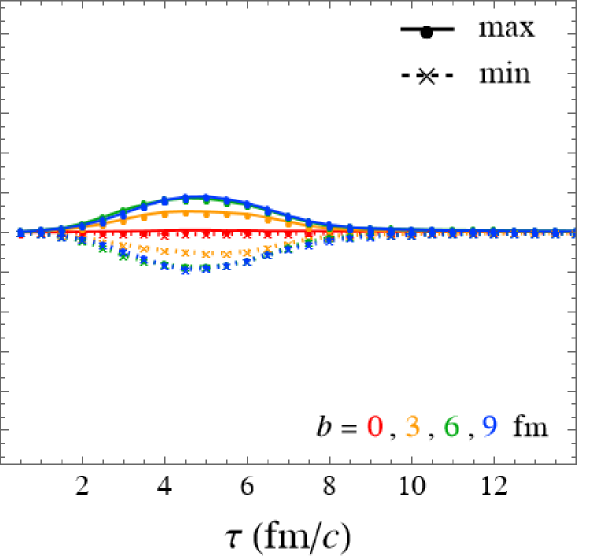}\hspace*{-2.35mm}
\includegraphics[align=t, height=0.22\textwidth, clip, trim = 139 0 15 42.5]{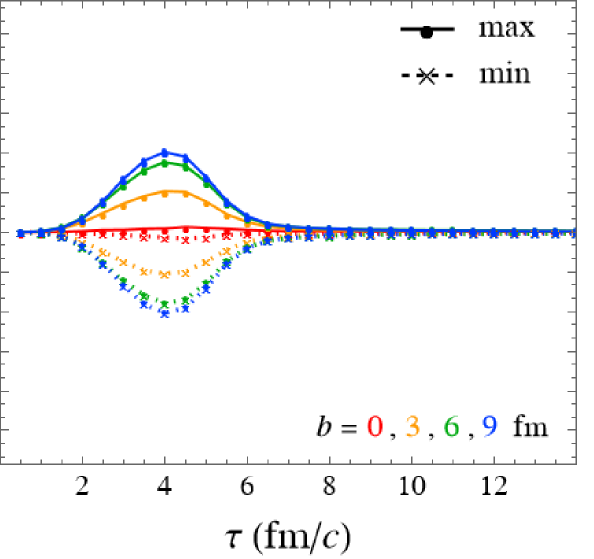}\hspace*{-2.35mm}
\includegraphics[align=t, height=0.22\textwidth, clip, trim = 139 0 15 42.5]{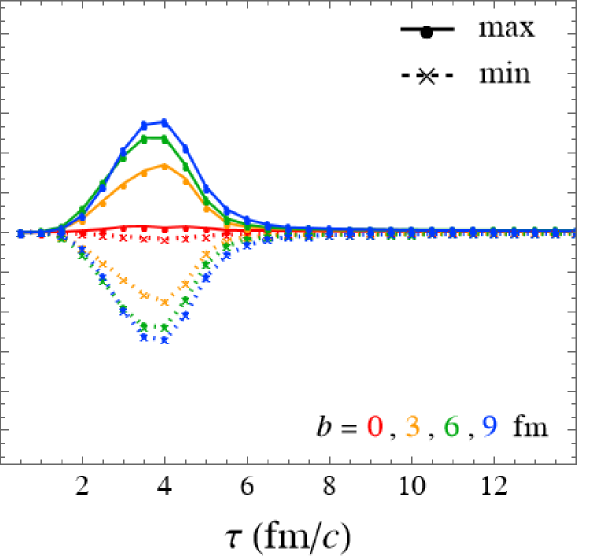}
\caption{\label{fig:4} The time evolution of the maximum (solid) and minimum (dashed) over the space of the Lorentz invariants, ${\mathcal F} = {\bm E}^2-{\bm B}^2$ (top) and ${\mathcal G}={\bm E}\cdot{\bm B}$ (bottom), for various collision energies $\sqsNN$ and impact parameters $b$.  }
\end{figure*}

I turn to discuss in more detail the time evolution of the produced electromagnetic field.  To this end, I compute the maximum and minimum values of the Lorentz invariants, ${\mathcal F}$ and ${\mathcal G}$, over space at each time step.  The results are presented in Fig.~\ref{fig:4}, where four representative collision energies, $\sqsNN = 3.0, 4.5, 6.2$, and $7.2\;{\rm GeV}$, are selected for illustration.

Before proceeding, I recall that maximizing (minimizing) the Lorentz invariant ${\mathcal F} = {\bm E}^2 - {\bm B}^2$ requires a large electric field ${\bm E}$ (magnetic field ${\bm B}$).  In other words, one can interpret $\max {\mathcal F} \approx \max {\bm E}^2$ and $\min {\mathcal F} \approx -\max {\bm B}^2$ as approximate measures of the maximum magnitudes of the electric and magnetic fields in space, respectively.

The top panel of Fig.~\ref{fig:4} shows the time evolution of $\max {\mathcal F}$ (solid) and $\min {\mathcal F}$ (dashed).  At the central collision, $b=0$, $\max {\mathcal F}$ is positive all over the space, while $\min {\mathcal F}$ is zero for all collision energies.  With increasing the impact parameter $b>0$, both $\max {\mathcal F}$ and $\min {\mathcal F}$ decrease.  In other words, the electric field is the strongest at the central collision and becomes weaker, with developing magnetic field, by going to non-central collisions.  This is a reasonable behavior, as anticipated in Introduction.  One also notices that, with increasing impact parameter, the magnitude of $\min {\mathcal F}$ grows gradually and eventually overwhelms that of $\max {\mathcal F}$ around $b=6\;{\rm fm}$.  This means that the magnetic field can be stronger than the electric field only for relatively large impact parameters $b\gtrsim 6\;{\rm fm}$.  Note that, even for $b\gtrsim 6\;{\rm fm}$ where the magnetic field becomes stronger than the electric field, the system is not necessarily dominated by the magnetic field and the electric field is still important because the spacetime volume of the magnetic field stays much smaller than that of the electric field (see also discussions in Figs.~\ref{fig:1} and \ref{fig:6}).  

The times of the peaks of $\max {\mathcal F}$ and $\min {\mathcal F}$ indicate when the electric and magnetic fields become the strongest.  The results show that both fields become the strongest simultaneously at the moment of a collision, i.e., when the colliding ions maximally overlap with each other and the charge density becomes the highest.  After the collision, as the charge diffuses, the electromagnetic field decays.  Although the decay rate increases with the collision energy, it remains relatively slow in the intermediate-energy regime due to the baryon stopping.  Consequently, the produced electromagnetic field is much more long-lived, ${\mathcal O}(1\,\mathchar`-\,10\;{\rm fm}/c)$, compared to high-energy collisions.  Note that $t=0$ is not defined to be the time of a collision but is just a starting time of a JAM simulation.  The starting time of JAM is defined in the default setting as the time when the longitudinal distance between the colliding ions becomes $z_0=3\;{\rm fm}$.  This means that the centers of the colliding ions at $t=0$ are $z = \pm(z_0/2 + R/\gamma)$, where $R \approx 6.4\;{\rm fm}$ is the radius of the colliding gold ions and $\gamma$ is the beam gamma factor, and hence the time of a collision can roughly be estimated as $t_{\rm coll} = (1.5 + 12.8 m_N/\sqsNN)\;{\rm fm}/c$, where $m_N \approx 0.94\;{\rm GeV}$ is the nucleon mass.  

One can make similar observations for $\max {\mathcal G}$ and $\min {\mathcal G}$, which are shown in the bottom panel of Fig.~\ref{fig:4}.  The magnitudes of $\max {\mathcal G}$ and $\min {\mathcal G}$ are also maximized at the instant of a collision, and then slowly decay with lifetime ${\mathcal O}(1\,\mathchar`-\,10\;{\rm fm}/c)$.

\subsection{Peak field strength}

\begin{figure*}[!t]
\hspace*{-90mm}
\includegraphics[align=t, width=0.49\textwidth, clip]{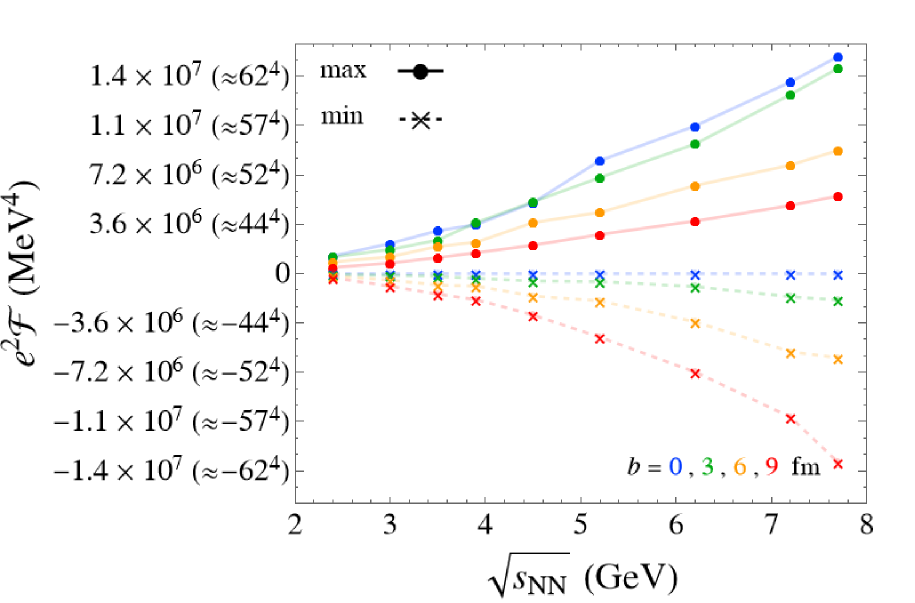}\hspace*{5mm}
\includegraphics[align=t, width=0.49\textwidth, clip]{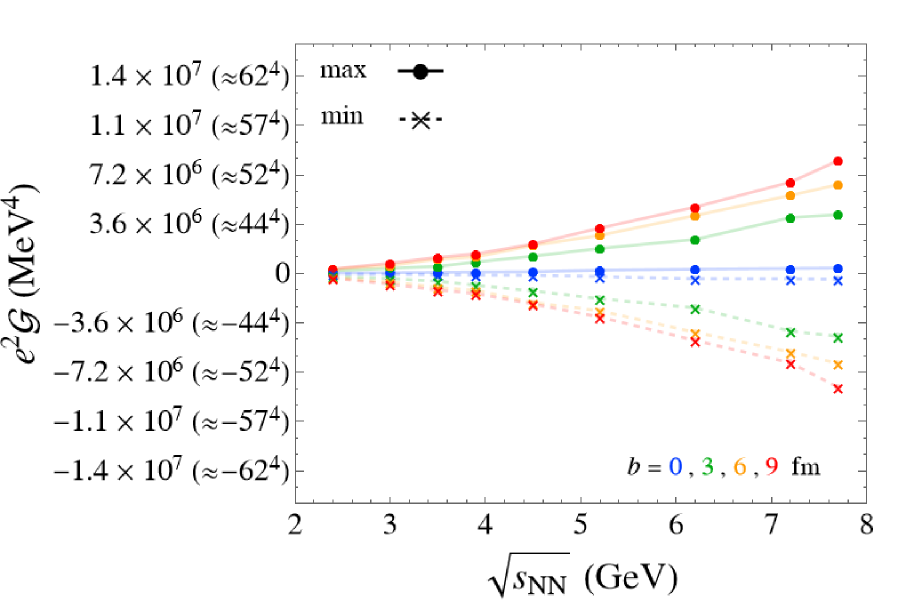}
\caption{\label{fig:5} The collision-energy $\sqsNN$ dependence of the maximum (solid) and minimum (dashed) values of the Lorentz invariants, ${\mathcal F} = {\bm E}^2-{\bm B}^2$ (left) and ${\mathcal G} = {\bm E}\cdot{\bm B}$ (right), for various impact parameters $b$.  }
\end{figure*}

To develop a quantitative insight, I extract the maximum and minimum values of the Lorentz invariants, ${\mathcal F}$ and ${\mathcal G}$, over the whole spacetime, and plot them as a function of the collision energy for various values of the impact parameter in Fig.~\ref{fig:5}.  

Figure~\ref{fig:5} indicates that the typical magnitudes of the maxima and minima of ${\mathcal F}$ and ${\mathcal G}$ are of the order of ${\mathcal O}((50\;{\rm MeV})^4)$.  This means that the typical magnitudes of the produced electromagnetic fields, ${\bm E}$ and ${\bm B}$, are of the order of ${\mathcal O}((50\;{\rm MeV})^2)$.  This scale, ${\mathcal O}(50\;{\rm MeV})^2$, is strong in a sense that it goes beyond the critical field strength of QED, $m_e^2 \approx (0.5\;{\rm MeV})^2$, and is non-negligibly large even compared to the QCD/hadronic scale $m_\pi \approx (140\;{\rm MeV})^2$, where $m_\pi$ is the pion mass.  

The magnitudes of the maxima and minima of ${\mathcal F}$ and ${\mathcal G}$ are monotonically increasing with the collision energy.  In other words, both electric and magnetic fields become stronger, as the collision energy increases.  This is essentially because the charge density becomes higher with larger collision energies due to the Lorentz contraction.  Note that it does not necessarily mean that high-energy collisions are more advantageous for the observations of strong-field effects: one must also care about the spacetime volume because effects may be negligible if the spacetime volume is tiny, no matter how strong the fields are~\cite{Taya:2024wrm}.  I will come back to this issue of spacetime volume in the next subsection.  

At the quantitative level, I find numerically that $\max {\mathcal F}$ can be fit well with a simple function of the form, ${\rm const}. \times (\sqsNN )^2$, as
\begin{subequations}
\begin{align}
	\max e^2{\mathcal F}(b=0\;{\rm fm}) 
		&\approx (23\;{\rm MeV})^4 \times \left( \frac{\sqsNN}{1\;{\rm GeV}}  \right)^2 \;, \\
	\max e^2{\mathcal F}(b=3\;{\rm fm}) 
		&\approx (22\;{\rm MeV})^4 \times \left( \frac{\sqsNN}{1\;{\rm GeV}}  \right)^2 \;, \\
	\max e^2{\mathcal F}(b=6\;{\rm fm}) 
		&\approx (20\;{\rm MeV})^4 \times \left( \frac{\sqsNN}{1\;{\rm GeV}}  \right)^2 \;, \\
	\max e^2{\mathcal F}(b=9\;{\rm fm}) 
		&\approx (18\;{\rm MeV})^4 \times \left( \frac{\sqsNN}{1\;{\rm GeV}}  \right)^2 \;.
\end{align}
\end{subequations}
For $\min {\mathcal F}, \max {\mathcal G}$, and $\min {\mathcal G}$, I find that the simple function, ${\rm const}. \times (\sqsNN )^2$, is not sufficient to fit the data but the addition of a constant term makes it working well, ${\rm const}. + {\rm const}. \times (\sqsNN )^2$.  I find
\begin{subequations}
\begin{align}
	&\min e^2{\mathcal F}(b=3\;{\rm fm}) \nonumber\\
		&\quad\approx (21\;{\rm MeV})^4 - (14\;{\rm MeV})^4 \times \left( \frac{\sqsNN}{1\;{\rm GeV}}  \right)^2\;, \\
	&\min e^2{\mathcal F}(b=6\;{\rm fm}) \nonumber\\
		&\quad\approx (29\;{\rm MeV})^4 - (18\;{\rm MeV})^4 \times \left( \frac{\sqsNN}{1\;{\rm GeV}}  \right)^2 \;, \\
	&\min e^2{\mathcal F}(b=9\;{\rm fm}) \nonumber\\ 
		&\quad\approx (35\;{\rm MeV})^4 - (22\;{\rm MeV})^4 \times \left( \frac{\sqsNN}{1\;{\rm GeV}}  \right)^2 \;.
\end{align}
\end{subequations}
and
\begin{subequations}
\begin{align}
	&\max e^2{\mathcal G}(b=3\;{\rm fm}) \approx - \min e^2{\mathcal G}(b=3\;{\rm fm}) \nonumber\\
		&\quad \approx (-25\;{\rm MeV})^4 + (17\;{\rm MeV})^4 \times \left( \frac{\sqsNN}{1\;{\rm GeV}}  \right)^2\;, \\
	&\max e^2{\mathcal G}(b=6\;{\rm fm}) \approx - \min e^2{\mathcal G}(b=6\;{\rm fm}) \nonumber\\
		&\quad \approx (-25\;{\rm MeV})^4 + (19\;{\rm MeV})^4 \times \left( \frac{\sqsNN}{1\;{\rm GeV}}  \right)^2 \;, \\
	&\max e^2{\mathcal G}(b=9\;{\rm fm}) \approx - \min e^2{\mathcal G}(b=9\;{\rm fm}) \nonumber\\
		&\quad \approx (-28\;{\rm MeV})^4 + (20\;{\rm MeV})^4 \times \left( \frac{\sqsNN}{1\;{\rm GeV}}  \right)^2 \;,
\end{align}
\end{subequations}
where $b=0\;{\rm fm}$ is omitted, for which $\min {\mathcal F}, \max {\mathcal G}$, and $\min {\mathcal G}$ are simply vanishing within the residual event-by-event fluctuations.

\subsection{Spacetime volume}

\begin{figure}[!t]
\hspace*{-70mm}\includegraphics[align=t, width=0.38\textwidth, clip]{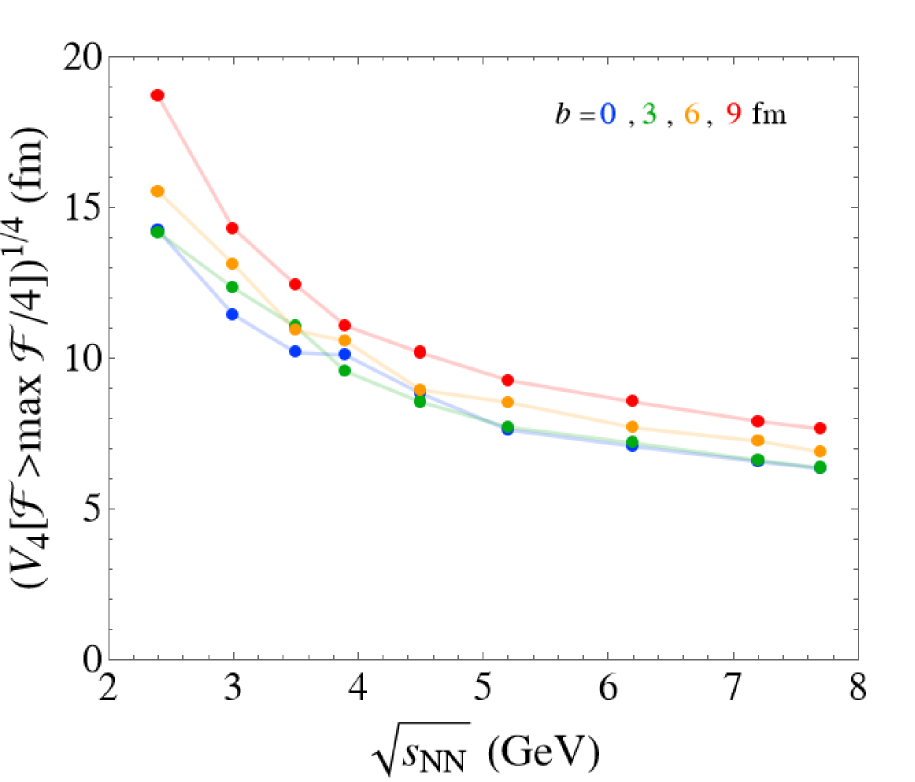}\\
\hspace*{-70mm}\includegraphics[align=t, width=0.38\textwidth, clip]{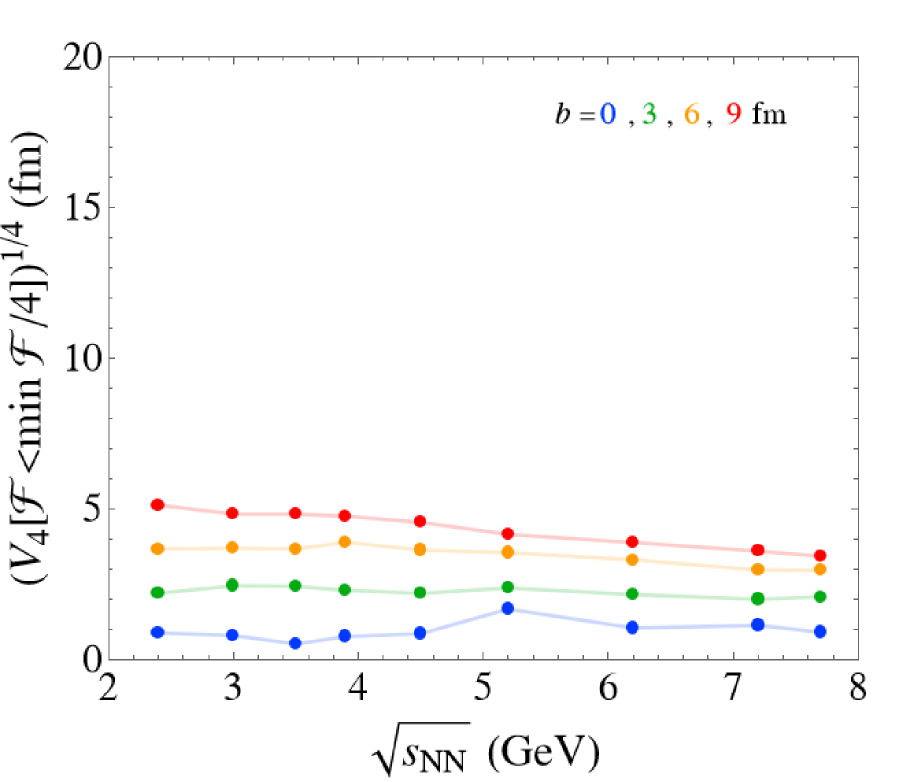}
\caption{\label{fig:6} The spacetime four-volume $V_4$ of the Lorentz invariant ${\mathcal F}={\bm E}^2-{\bm B}^2$, defined as the region that satisfies ${\mathcal F} > \max {\mathcal F}/4$ (top) and ${\mathcal F} < \min {\mathcal F}/4$ (bottom).    }
\end{figure}

\begin{figure}[!t]
\hspace*{-70mm}\includegraphics[align=t, width=0.38\textwidth, clip]{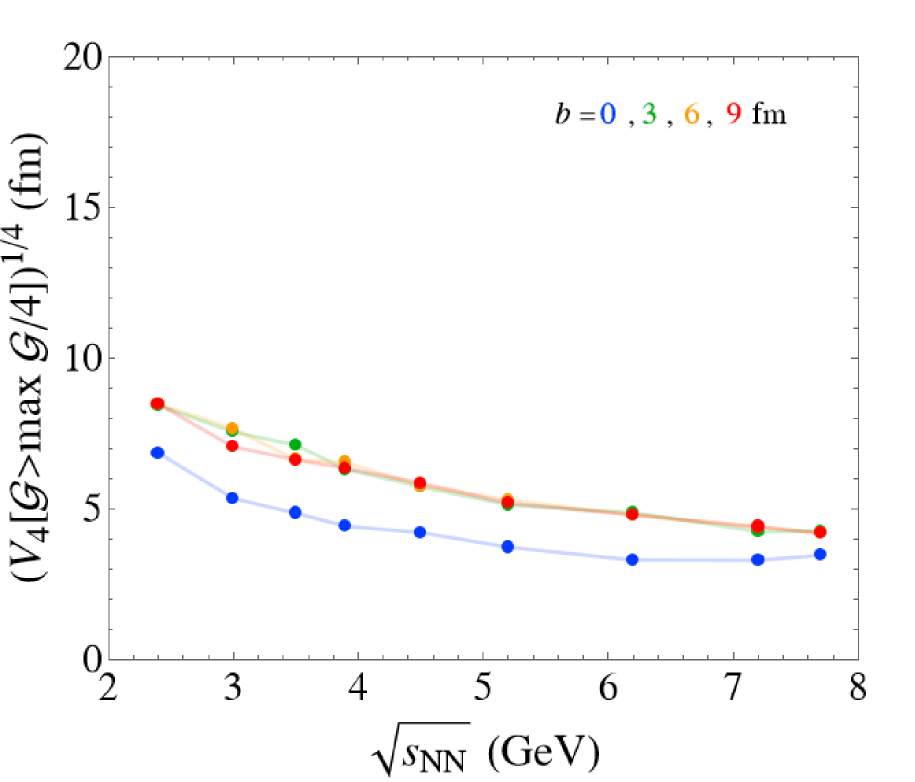}\\
\hspace*{-70mm}\includegraphics[align=t, width=0.38\textwidth, clip]{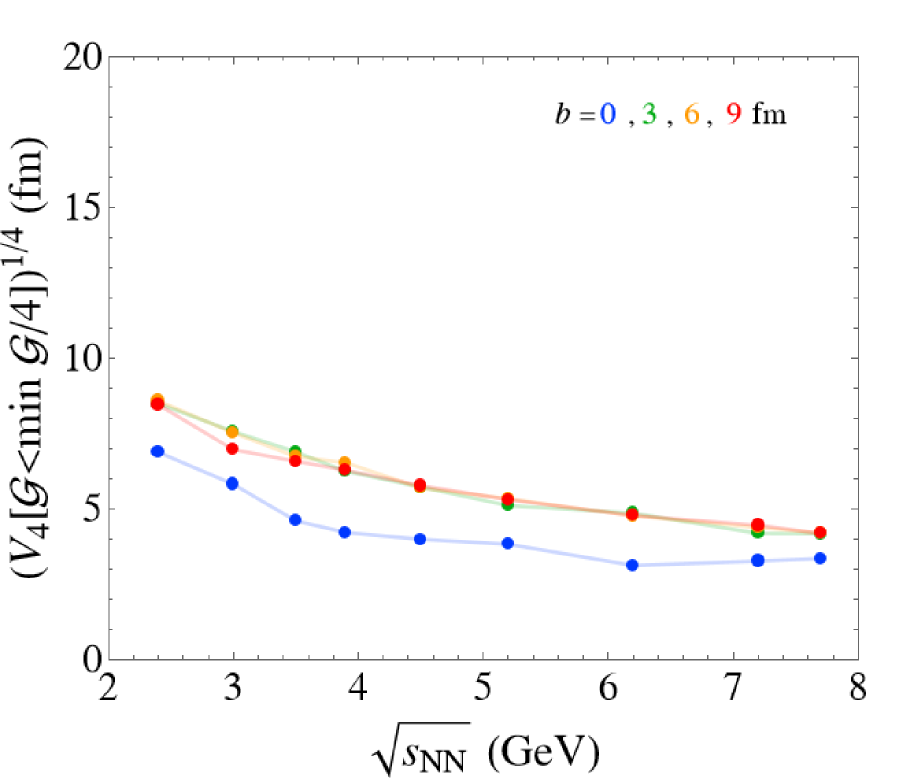}
\caption{\label{fig:7} A plot similar to Fig.~\ref{fig:6}, showing the spacetime four-volume $V_4$ of the Lorentz invariant ${\mathcal G}={\bm E}\cdot{\bm B}$, defined as the region such that ${\mathcal G} > \max {\mathcal G}/4$ (top) and ${\mathcal G} < \min {\mathcal G}/4$ (bottom).    }
\end{figure}

Finally, I quantitatively discuss the spacetime volume of the produced electromagnetic field.  I quantify the spacetime volume $V_4$ by the region where the Lorentz invariants satisfy the following threshold conditions: 
\begin{subequations}
\begin{align}
	V_4[{\mathcal F}>\max {\mathcal F}/4] := \int_{ {\mathcal F}>\max {\mathcal F}/4 }{\rm d}^4x \;,\\
	V_4[{\mathcal F}<\min {\mathcal F}/4] := \int_{ {\mathcal F}<\min {\mathcal F}/4 }{\rm d}^4x \;,\\
	V_4[{\mathcal G}>\max {\mathcal G}/4] := \int_{ {\mathcal G}>\max {\mathcal F}/4 }{\rm d}^4x \;,\\
	V_4[{\mathcal G}<\min {\mathcal G}/4] := \int_{ {\mathcal G}<\min {\mathcal F}/4 }{\rm d}^4x \;,
\end{align}
\end{subequations}
where $\max$ and $\min$ mean to take the maximum and minimum over the whole spacetime, respectively.  The physical meanings of these volumes are evident.  For example, $V_4[{\mathcal F}>\max {\mathcal F}/4]$ and $V_4[{\mathcal F}<\min {\mathcal F}/4]$, respectively, measure the volumes of the strong electric- and magnetic-field regions such that ${\mathcal F} > \max {\mathcal F}/4 \approx \max {\bm E}^2/4$ and ${\mathcal F} < \min {\mathcal F}/4 \approx -\max {\bm B}^2/4$.  Note that ${\mathcal F}$ and ${\mathcal G}$ are roughly the squares of the electromagnetic fields, ${\bm E}$ and ${\bm B}$, and hence the denominators ``4" amount to considering the half maxima of ${\bm E}$ and ${\bm B}$.  I have verified that the qualitative features of the spacetime volume remain unchanged for other values of denominators, such as 2 and 6.  

The results are shown in Fig.~\ref{fig:6} for ${\mathcal F}$ and in Fig.~\ref{fig:7} for ${\mathcal G}$.  One first observes that the typical magnitude of $V_4$ is much larger in the top panel of Fig.~\ref{fig:6} compared to the other three panels, i.e., $V_4[{\mathcal F}>\max {\mathcal F}/4] \gg V_4[({\rm others})]$.  This reflects the fact that the electric field occupies majority of the spacetime, as shown in Fig.~\ref{fig:1}.  It is remarkable that even at the most non-central case $b=9\;{\rm fm}$, one finds $V_4[{\mathcal F}>\max {\mathcal F}/4] \approx (10\;{\rm fm})^4 \gg V_4[{\mathcal F}<\min {\mathcal F}/4] \approx (5\;{\rm fm})^4$.  In other words, the region dominated by the magnetic field $V_4[{\mathcal F}<\min {\mathcal F}/4]$ is always smaller than that dominated by the electric field $V_4[{\mathcal F}>\max {\mathcal F}/4]$ in intermediate-energy heavy-ion collisions.

Among the four $V_4$'s, one finds that $V_4[{\mathcal F}>\max {\mathcal F}/4]$ has the strongest collision-energy dependence.  It scales with $1/\sqsNN$ due to the Lorentz contraction.  Quantitatively, $V_4[{\mathcal F}>\max {\mathcal F}/4]$ can be fit well with a function of the form ${\rm const}. + {\rm const.}/\sqsNN$.  The numerical coefficients are found to be 
\begin{subequations}
\begin{align}
	&V_4[{\mathcal F}>\max {\mathcal F}/4](b=0\;{\rm fm}) \nonumber\\
		&\quad\approx (2.7\;{\rm fm}) + (27\;{\rm fm}) \times \left( \frac{\sqsNN}{1\;{\rm GeV}}  \right)^{-1}\;, \\
	&V_4[{\mathcal F}>\max {\mathcal F}/4](b=3\;{\rm fm}) \nonumber\\
		&\quad\approx (2.5\;{\rm fm}) + (28\;{\rm fm}) \times \left( \frac{\sqsNN}{1\;{\rm GeV}}  \right)^{-1}\;, \\
	&V_4[{\mathcal F}>\max {\mathcal F}/4](b=6\;{\rm fm}) \nonumber\\
		&\quad\approx (2.8\;{\rm fm}) + (30\;{\rm fm}) \times \left( \frac{\sqsNN}{1\;{\rm GeV}}  \right)^{-1}\;, \\
	&V_4[{\mathcal F}>\max {\mathcal F}/4](b=9\;{\rm fm}) \nonumber\\
		&\quad\approx (2.3\;{\rm fm}) + (37\;{\rm fm}) \times \left( \frac{\sqsNN}{1\;{\rm GeV}}  \right)^{-1}\;. 
\end{align}
\end{subequations}
I apply the same function to fit the other $V_4$'s, finding (omitting the uninterested $b=0$ case)
\begin{subequations}
\begin{align}
	&V_4[{\mathcal F}<\min {\mathcal F}/4](b=3\;{\rm fm}) \nonumber\\
		&\quad\approx (2.0\;{\rm fm}) + (0.88\;{\rm fm}) \times \left( \frac{\sqsNN}{1\;{\rm GeV}}  \right)^{-1}\;, \\
	&V_4[{\mathcal F}<\min {\mathcal F}/4](b=6\;{\rm fm}) \nonumber\\
		&\quad\approx (2.9\;{\rm fm}) + (2.5\;{\rm fm}) \times \left( \frac{\sqsNN}{1\;{\rm GeV}}  \right)^{-1}\;, \\
	&V_4[{\mathcal F}<\min {\mathcal F}/4](b=9\;{\rm fm}) \nonumber\\
		&\quad\approx (2.3\;{\rm fm}) + (5.9\;{\rm fm}) \times \left( \frac{\sqsNN}{1\;{\rm GeV}}  \right)^{-1}\;,
\end{align}
\end{subequations}
and 
\begin{subequations} \label{eq:14}
\begin{align}
	&V_4[{\mathcal G}>\max {\mathcal G}/4](b=3\;{\rm fm}) \approx V_4[{\mathcal G}<\min {\mathcal G}/4](b=3\;{\rm fm}) \nonumber\\
		&\quad\approx (2.2\;{\rm fm}) + (15\;{\rm fm}) \times \left( \frac{\sqsNN}{1\;{\rm GeV}}  \right)^{-1}\;, \\
	&V_4[{\mathcal G}>\max {\mathcal G}/4](b=6\;{\rm fm}) \approx V_4[{\mathcal G}<\min {\mathcal G}/4](b=6\;{\rm fm}) \nonumber\\
		&\quad\approx (2.3\;{\rm fm}) + (15\;{\rm fm}) \times \left( \frac{\sqsNN}{1\;{\rm GeV}}  \right)^{-1}\;, \\
	&V_4[{\mathcal G}>\max {\mathcal G}/4](b=9\;{\rm fm}) \approx V_4[{\mathcal G}<\min {\mathcal G}/4](b=9\;{\rm fm}) \nonumber\\
		&\quad\approx (2.4\;{\rm fm}) + (15\;{\rm fm}) \times \left( \frac{\sqsNN}{1\;{\rm GeV}}  \right)^{-1}\;.
\end{align}
\end{subequations}

It is interesting to observe that the spacetime volume $V_4$ of ${\mathcal G}$ appears to be largely insensitive to the impact parameter.  One might interpret this insensitivity as a reflection of the dominance of the Coulomb electric field in intermediate-energy heavy-ion collisions and its long-range nature.  Indeed, the spacetime extents of the electric field (see Fig.~\ref{fig:1}) and the finite ${\mathcal G}$ region (see Fig.~\ref{fig:3}) are roughly of comparable magnitude and significantly exceed the overlap region of the colliding nuclei, where the magnetic field can be effective.  This observation implies that the spacetime volume $V_4$ of ${\mathcal G}$ is primarily determined by the electric field\footnote{To be strict, the dependence on the impact parameter is, of course, not dictated solely by the electric field.  The magnetic field should also contribute because ${\mathcal G}$ contains ${\bm B}$, and the competition between the electric and magnetic fields finally determines the spacetime volume.  However, it involves the complex dynamics of heavy-ion reactions, and, at present, I lack a simple explanation for this.  }.  Since the Coulomb electric field is long-ranged, its associated spacetime volume $V_4$ becomes relatively insensitive to variations in the impact parameter $b$, as observed in Fig.~\ref{fig:6}.  Consequently, the $V_4$ of ${\mathcal G}$ inherits this insensitivity. 

\section{Summary and discussion} \label{sec:4}

I have numerically estimated the electromagnetic field produced in intermediate-energy heavy-ion collisions with finite impact parameter by using a hadron transport model, JAM.  

The main findings are (1) that the produced electromagnetic field is strong $eE, eB = {\mathcal O}((50\;{\rm MeV})^2)$, which goes beyond the critical field strength of QED $m_e^2 \approx (0.5\;{\rm MeV})^2$ and is non-negligibly large even compared to the QCD/hadronic scale $m_\pi^2 \approx (140\;{\rm MeV})^2$; (2) that the produced fields extend in spacetime about $V_4 = {\mathcal O}((10\;{\rm fm})^4)$, which is larger than the typical spacetime volume of the fireball created in heavy-ion collisions~\cite{Taya:2024zpv} and is also large in terms of the nonlinearity parameters of QED~\cite{Taya:2014taa, Taya:2024wrm, Siddique:2025tzd}; (3) that the majority of the spacetime is occupied by the electric field and the magnetic field dominates only around the collision point; and (4) that a topological electromagnetic-field configuration such that ${\mathcal G} = {\bm E}\cdot {\bm B} \neq 0$ is realized, which can induce nontrivial chiral phenomena through the Adler-Bell-Jackiw (ABJ) chiral anomaly~\cite{Adler:1969gk, Bell:1969ts}.  

The present results represent a first step toward quantifying electromagnetic-field effects in intermediate-energy heavy-ion collisions.  The next step is to use the electromagnetic-field profile obtained in this study as input for predicting observable signatures, such as the dilepton production and the charged directed flow.  

I discuss a few more implications of the present work (cf. see also Ref.~\cite{Taya:2024wrm} for related discussions to strong-field QED and dense QCD).  While a quantitative investigation of these implications and their observable consequences lie beyond the scope of this work (as emphasized in the last paragraph, the present study aims to serve as an initial step toward it), I hope to stimulate further discussions and exploration in the community.  
\begin{itemize}
\item {\it Chiral phenomena}.--- The finite ${\bm E}\cdot {\bm B}$ can induce nontrivial chiral-anomaly driven phenomena, including the chirality production~\cite{Fukushima:2010vw, Warringa:2012bq, Copinger:2018ftr, Taya:2020bcd, Fukushima:2023obj} and the chiral plasma instability~\cite{Akamatsu:2013pjd}.  Among these, I highlight the implication to the anomaly-induced transport phenomena such as the chiral-magnetic effect and the chiral-vortical effect~(see Ref.~\cite{Kharzeev:2024zzm} for the latest review).  

The present results suggest that intermediate-energy heavy-ion collisions provide a unique environment for the experimental search for anomaly-induced transport phenomena, while the high-energy regime has been the primary focus of such searches over the past decade and there exist only a few experimental results in the intermediate-energy regime~\cite{STAR:2014uiw, STAR:2022ahj, Xu:2023wcy}.  For the anomaly-induced transport phenomena to occur, one must have a chirality imbalance, or a finite chiral chemical potential $\mu_5$, and an external field (such as magnetic field ${\bm B}$ and vorticity ${\bm \omega}$) that couples to $\mu_5$ and drives a current.  In intermediate-energy heavy-ion collisions, the chirality imbalance is naturally provided by the topological electromagnetic-field configuration through the ABJ anomaly as $\mu_5 \propto {\bm E}\cdot {\bm B} \neq 0$ (see Fig.~\ref{fig:3}).  One also has a strong magnetic field, as revealed in the present paper (see Figs.~\ref{fig:1} and \ref{fig:6}), and a strong vorticity, as shown by other authors~\cite{Deng:2020ygd}.  Therefore, intermediate-energy heavy-ion collisions do satisfy the conditions for the anomaly-induced transport phenomena to occur.  Consequently, one can expect transports of electric charge $\propto \mu_5 {\bm B}$ via the chiral-magnetic effect and baryon charge $\propto \mu_5 \mu {\bm \omega}$ via the chiral-vortical effect, with $\mu$ being the vector chemical potential.  Note that other types of the anomaly-induced transport phenomena may also occur in principle, e.g., the chiral-electric-separation effect~\cite{Huang:2013iia, Jiang:2014ura} driven by the strong electric field.  

In the search for anomaly-induced transport phenomena in intermediate-energy heavy-ion collisions, it is important to recognize some differences compared to the high-energy case.  First, the source of the chirality imbalance $\mu_5$ differs: in the intermediate-energy case, it arises from the topological configuration of the electromagnetic field, whereas in high-energy collisions, it is generated by the so-called Glasma~\cite{Lappi:2006fp}, i.e., color flux tubes that span between the colliding ions.  Owing to these differing origins, the spacetime profiles of the resulting $\mu_5$ are distinct, which should be reflected in observable signals.  For example, the Glasma, or the associated $\mu_5$, is randomly distributed in the transverse plane and thus is vanishing upon the event averaging; in contrast, the topological electromagnetic-field configuration has a characteristic spatial structure, such as the sign flipping in the upper and lower half planes, and survives even after the event averaging (see Fig.~\ref{fig:3}).  Second, as the topological charge is sourced by the electromagnetic field, it may result in anomaly-induced transport phenomena not only of quarks/hadrons but also of leptons such as electrons and muons, which do not couple to the color field (i.e., the Glasma) and hence are not driven by the anomaly-induced transport at high energies.  Exploring such leptonic contributions and their experimental signatures represents an intriguing direction for future study.  Third, intermediate-energy collisions offer a potential advantage over high-energy ones: the magnetic field ${\bm B}$ (see Fig.~\ref{fig:6}) and vorticity ${\bm \omega}$~\cite{Deng:2020ygd} persist longer in time and thus may enhance the signals.  

\item {\it Importance of the electric field}.---  As demonstrated in the present study, the electric field exhibits a larger spacetime volume and magnitude than the magnetic field, and thus can exert a greater influence.  This finding may represent a significant shift in the study of electromagnetic-field effects in heavy-ion collisions, where prior research has predominantly focused on the magnetic field (see Ref.~\cite{Hattori:2016emy} for a review).

Electric fields influence the collision system in a manner distinct from magnetic fields and can therefore leave unique signatures.

For example, it is expected that electric and magnetic fields lead to opposite signs in the charge-dependent directed flow, defined as $\Delta v_1 := v_1({\rm positive\ charge}) - v_1({\rm negative\ charge})$.  Specifically, $\Delta v_1$ is predicted to be negative for electric fields and positive for magnetic fields in the forward rapidity region (with the signs reversed in the backward region) for electric and magnetic fields, respectively~\cite{Gursoy:2014aka}.  Interestingly, a recent experimental result at an intermediate energy of $\sqsNN = 27\;{\rm GeV}$ reported a trend in $\Delta v_1$ consistent with the electric-field dominance~\cite{STAR:2023jdd}, potentially lending experimental support to the findings of the present paper.  

Another example is that while electric fields supply energy to the system,  magnetic fields do not.  This difference leads to intriguing electric phenomena such as the (Sauter-)Schwinger effect, i.e., the spontaneous pair production of charged particles from the vacuum~\cite{Sauter:1931zz, Schwinger:1951nm}.  It is one of the goals of strong-field QED to experimentally verify the Schwinger effect.  Therefore, it is a very interesting possibility to use intermediate-energy heavy-ion collisions as a new tool to achieve this.  As a potential experimental signature, I naively expect an excess of dileptons that cannot be accounted for by the conventional hadronic cocktails in the very-low momentum regime of the order of the strength of the electric field $\approx 50 \;{\rm MeV}$, according to the celebrated Schwinger formula~\cite{Schwinger:1951nm}.  However, the realistic electromagnetic-field configuration has a complicated spacetime structure, as shown in the present paper (see Figs~\ref{fig:1} and \ref{fig:3}).  This means that the naive estimate based on the Schwinger formula (or its slight extensions such as the locally-constant-field approximation~\cite{Bulanov:2004de, Aleksandrov:2018zso, Sevostyanov:2020dhs} and the semi-classical approaches~\cite{Brezin:1970xf, Dunne:2005sx, Taya:2020dco}) may be invalid, as it is applicable when the field is approximately constant in spacetime.  Developing a more sophisticated formula for the Schwinger effect is not only needed for a quantitative prediction of the signals based on the realistic field configuration I obtained in the present paper but also is a fundamental theoretical problem for the Schwinger effect, since its formulation for inhomogeneous fields has not been established (see Ref.~\cite{Fedotov:2022ely} for review).  The dense-matter effects may also contaminate the dilepton signals, and conversely those of the dense QCD can also be affected by the Schwinger effect, or strong-field-QED effects broadly speaking.  

As a final example, I mention the impacts on spin observables.  In heavy-ion collisions, the spin polarization of hadrons (e.g., the $\Lambda$ hyperon) has been measured experimentally~(see Ref.~\cite{Niida:2024ntm} for a recent review).  Theoretically, the spin polarization is primarily attributed to the strong vorticity of the produced matter and would also be affected by other effects such as the Zeeman effect driven by a magnetic field.  The present findings suggest the presence of a strong electric field and a topologically nontrivial electromagnetic-field configuration, which, respectively, can influence the spin polarization through the spin-orbit coupling and the chiral anomaly.  This presents an intriguing direction for further investigation, where spin-polarization measurements could, in turn, serve as experimental probes of the electromagnetic-field structure in heavy-ion collisions.

\item {\it Electric conductivity}.--- In this study, I have neglected the interaction between the produced field and the dense baryonic matter.  To obtain a more realistic estimate, such an interaction should also be included.  Notably, in high-energy contexts, it has been argued that the electric conductivity of the matter can prolong the lifetime of the magnetic field via Faraday induction~\cite{Tuchin:2010vs, McLerran:2013hla, Gursoy:2014aka, Tuchin:2015oka, Li:2016tel, Stewart:2021mjz, Benoit:2025amn}.  The electric conductivity also affects the lifetime of the electric field, which, however, would shorten its lifetime.  This is simply because the electric field is screened out by dissipating its energy by driving a current.  

To illustrate how the lifetimes of the electromagnetic fields can be modified, I consider a simplified (but frequently-considered) scenario with the following assumptions: (i) Ohm's law ${\bm J} = \sigma {\bm E}$, with $\sigma$ being the electric conductivity, holds; (ii) the electric conductivity $\sigma$ is just a constant; and (iii) the characteristic size of the field region $R$ is sufficiently large as $\sigma R^2 \gg 1$.  Under these assumptions, by simultaneously solving the Faraday and Amp\`{e}re laws, one can readily show ${\bm B} \propto {\rm e}^{- t/\sigma R^2}$~\cite{Tuchin:2010vs, Huang:2015oca, Hattori:2016emy}.  Similarly, from the Amp\`{e}re law, ${\bm E} \propto {\rm e}^{-\sigma t}$ follows immediately.  Thus, the magnetic field's lifetime is extended, while the electric field's is shortened.

However, I must emphasize that the assumptions are quite strong in the naive argument above: (i) Ohm's law in the form ${\bm J} = \sigma {\bm E}$ is not justified in the presence of a matter flow ${\bm v}$, which modifies the charge current as ${\bm J} \to \sigma ({\bm E} + {\bm v} \times {\bm B})$.  Thus, the dynamical evolution of the flow ${\bm v}$ also affects the lifetime, which can be significant.  Indeed, in the ideal magnetohydrodynamic limit where the electrical resistance vanishes, ${\bm E} = -{\bm v} \times {\bm B}$ holds and thus the electric field is not screened but instead co-evolves with the magnetic field (whose lifetime stays long in this limit~\cite{Huang:2015oca, Hattori:2016emy}).  (ii) The assumption of a constant $\sigma$ implicitly presumes weak fields and negligible spacetime variations of matter properties such as temperature and density.  In general, $\sigma$ is a function of all of these quantities, whose magnitudes are typically large in relativistic heavy-ion collisions, necessitating modification of the naive argument.  Quantifying the parameter dependencies of $\sigma$, particularly in intermediate-energy collisions, remains an important open theoretical problem.  Various studies have been done in the past decade to estimate the electric conductivity in the high-energy regime, where perturbative methods are applicable thanks to the high temperature and even the first-principle lattice calculation is feasible~\cite{Astrakhantsev:2019zkr, Aarts:2020dda, Almirante:2024lqn} because the baryon density is low and hence there is no sign problem.  The situation is drastically different in the intermediate-energy regime, where the temperature is not extremely high and a finite baryon density exists.  Besides, as I have shown in the present paper, not only a strong magnetic field but also a strong electric field exists.  Since they are strong, their impacts on the electric conductivity should be taken into account in a nonperturbative manner (cf. see Ref.~\cite{Taya:2023ltd} for a strong electric field at the zero temperature and density and Refs.~\cite{Li:2018ufq, Astrakhantsev:2019zkr, Ghosh:2024fkg, Almirante:2024lqn} for a strong magnetic field).  (iii) At the event-averaged level, it would be a reasonable assumption to regard the spatial variation of the electric field as small, or $R$ large, as it extends smoothly and widely in space according to the Coulomb law, while it may not be appropriate for the magnetic field as it is localized around the collision point (see Fig.~\ref{fig:1}).  Meanwhile, as I shall discuss soon, event-by-event fluctuations of the electromagnetic fields are not small in general, casting a doubt on the assumption of large $R$.  This can make a significant difference.  As a demonstration, consider the extreme limit $R \to 0$, for which the fields behave like on-shell photons and evolve according to the wave equations in the vacuum, $0 = (\partial_t^2 - \nabla^2){\bm E} = (\partial_t^2 - \nabla^2){\bm B}$.  Since these equations are independent of the matter properties, they imply in such a limit of $R \to 0$ that the fields are no longer affected by the matter and simply propagate at the speed of light.

Note that, even if the electric field were rapidly screened due to the matter effect, this does not render it irrelevant to heavy-ion collisions.  Indeed, the screening process converts the field energy into the matter.  The energy density of the field can be substantial, on the order of ${\bm E}^2 \approx {\mathcal O}((50\;{\rm MeV})^4/e^2) \approx {\mathcal O}((100\;{\rm MeV})^4)$, which is comparable to the typical energy-density of the matter, ${\mathcal O}(\Lambda_{\rm QCD}^4) \approx {\mathcal O}((200\;{\rm MeV})^4)$.  If the field energy is fully transferred to the matter, it may impart sizable momenta to charged particles constituting the matter, potentially leading to observable phenomena such as charge-dependent flow.

\item {\it Event-by-event fluctuations}.--- All the results presented in this work are obtained after the event averaging~(\ref{eq:5}).  It is simply because of the event averaging that the obtained electromagnetic-field profile turned out to be smooth in spacetime (see Figs.~\ref{fig:1} and \ref{fig:3}).  On an event-by-event basis, the phase-space distributions of the hadrons must fluctuate (due to, e.g., fluctuations of the initial positions of nucleons on the colliding ions and the probabilistic nature of collisions), resulting in local spacetime structures in the field (see Ref.~\cite{Deng:2012pc} for the high-energy case).  Accordingly, not only the magnitude of the electromagnetic field but also the spacetime volume must fluctuate event-by-event.  

It is an interesting direction to study the event-by-event fluctuations of the electromagnetic field.  To make a rough estimate of this, I mention Ref.~\cite{Taya:2024zpv}, in which event-by-event fluctuations of the baryon density in intermediate-energy heavy-ion collisions were studied.  It has shown that the spacetime volume of the baryon density can fluctuate about 20\;\%.  Since the baryon density is roughly twice the charge density and thus they would have a linear relation to each other, one can naively expect that the spacetime fluctuations of the electromagnetic field is also of the order of ${\mathcal O}(20\;\%)$.  This is relatively a large fluctuation, which needs to be tamed to catch rare signals originating from the strong electromagnetic field such as the Schwinger effect.  

Note that the insensitivity of the spacetime volume $V_4$ of ${\mathcal G}$ observed in Fig.~\ref{fig:7} is also obtained after the event averaging.  It is a-priori not trivial if the insensitivity persists on an event-by-event basis, and thus would be interesting to be pursued as future study.  Naively, the long-range nature of the Coulomb electric field should remain the same even on an event-by-event basis, and therefore I expect that the insensitivity still persists, if the reasoning provided below Eq.~(\ref{eq:14}) is correct.  
\end{itemize}

\section*{Acknowledgments}
The author thanks Toru~Nishimura and Akira~Ohnishi for the collaboration at the early stage of this work and Kensuke~Homma, Asanosuke~Jinno, Masakiyo~Kiatazawa, and Yasushi~Nara for enlightening discussions.  The author also thanks the West Lake Workshop on Nuclear Physics 2024, where these results were reported and the author had fruitful discussions with the participants.  This work was supported by JSPS KAKENHI under Grant No.~24K17058 and the RIKEN TRIP initiative (RIKEN Quantum).

\appendix

\section{Spacetime profile of electric and magnetic fields} \label{sec:app1}

\begin{figure*}[!t]
\flushleft{\hspace*{15.5mm}\mbox{$t=1.0\;{\rm fm}/c$ \hspace{15.6mm} $4.5\;{\rm fm}/c$ \hspace{17.6mm} $8.0\;{\rm fm}/c$ \hspace{17.6mm} $11.5\;{\rm fm}/c$ \hspace{17mm} $15.0\;{\rm fm}/c$}} \\
\vspace*{1mm}
\hspace*{-33mm}
\includegraphics[align=t, height=0.18\textwidth, clip, trim = 35 43 105 32]{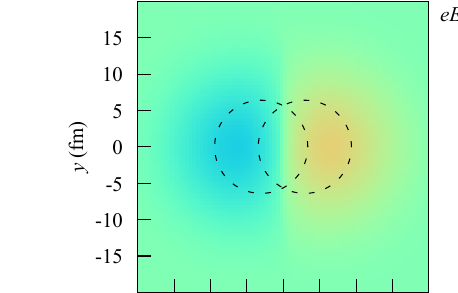}\hspace*{1.1mm}
\includegraphics[align=t, height=0.18\textwidth, clip, trim = 95 43 105 32]{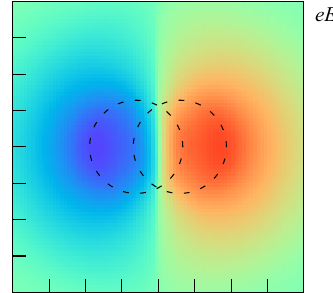}\hspace*{-5.7mm}
\includegraphics[align=t, height=0.18\textwidth, clip, trim = 95 43 105 32]{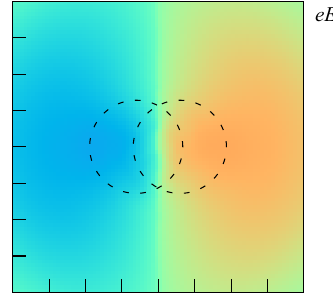}\hspace*{-5.7mm}
\includegraphics[align=t, height=0.18\textwidth, clip, trim = 95 43 105 32]{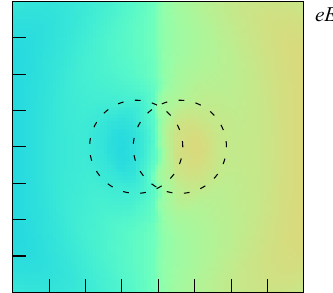}\hspace*{-11.4mm}
\includegraphics[align=t, height=0.18\textwidth, clip, trim = 95 43  55 32]{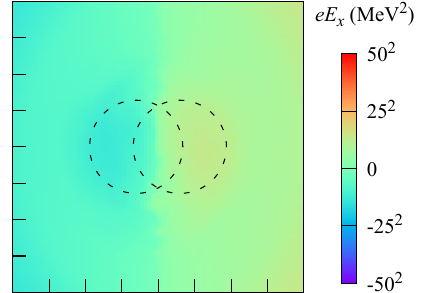}\hspace*{-6mm} \\
\vspace*{-0.7mm}
\hspace*{-33mm}
\includegraphics[align=t, height=0.18\textwidth, clip, trim = 35 43 105 32]{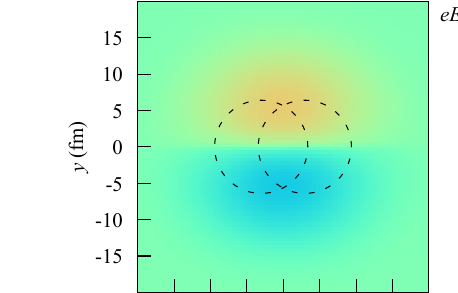}\hspace*{1.1mm}
\includegraphics[align=t, height=0.18\textwidth, clip, trim = 95 43 105 32]{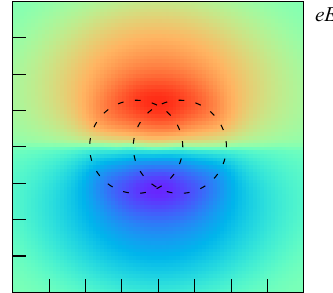}\hspace*{-5.7mm}
\includegraphics[align=t, height=0.18\textwidth, clip, trim = 95 43 105 32]{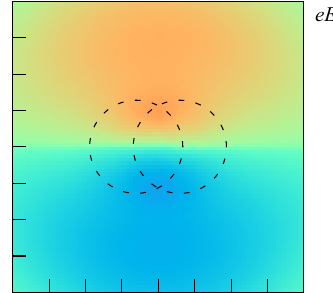}\hspace*{-5.7mm}
\includegraphics[align=t, height=0.18\textwidth, clip, trim = 95 43 105 32]{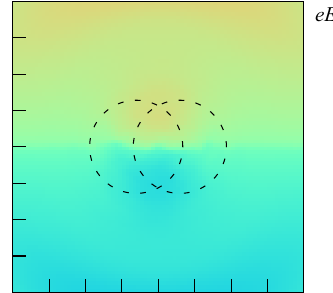}\hspace*{-11.4mm}
\includegraphics[align=t, height=0.18\textwidth, clip, trim = 95 43  55 32]{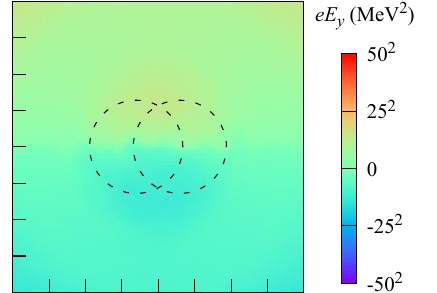}\hspace*{-6mm} \\
\vspace*{-0.7mm}
\hspace*{-33mm}
\includegraphics[align=t, height=0.18\textwidth, clip, trim = 35 43 105 32]{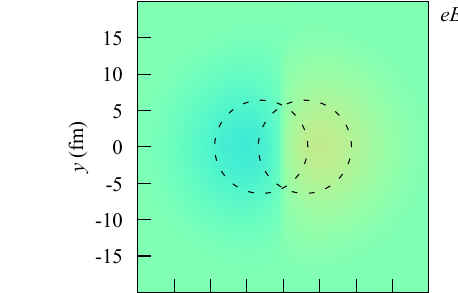}\hspace*{1.1mm}
\includegraphics[align=t, height=0.18\textwidth, clip, trim = 95 43 105 32]{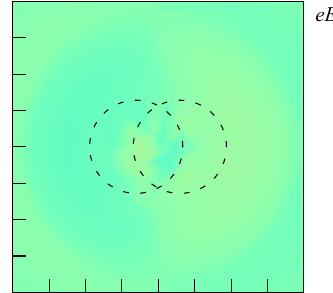}\hspace*{-5.7mm}
\includegraphics[align=t, height=0.18\textwidth, clip, trim = 95 43 105 32]{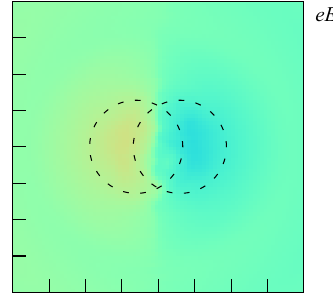}\hspace*{-5.7mm}
\includegraphics[align=t, height=0.18\textwidth, clip, trim = 95 43 105 32]{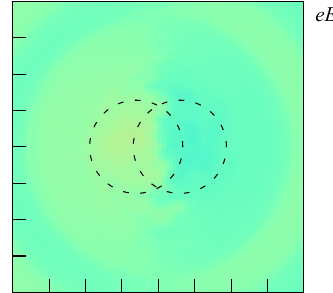}\hspace*{-11.4mm}
\includegraphics[align=t, height=0.18\textwidth, clip, trim = 95 43  55 32]{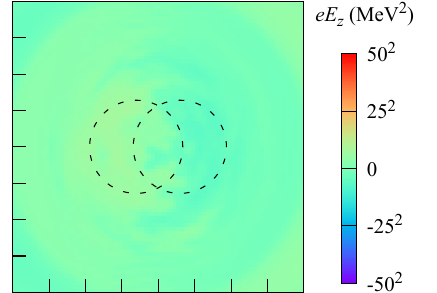}\hspace*{-6mm} \\
\vspace*{2mm}
\hspace*{-33mm}
\includegraphics[align=t, height=0.18\textwidth, clip, trim = 35 43 105 32]{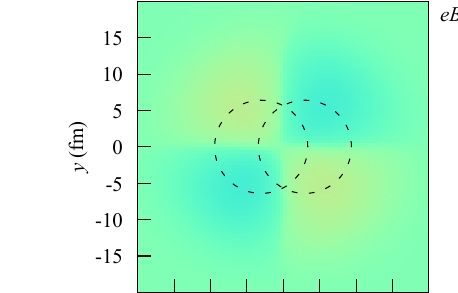}\hspace*{1.1mm}
\includegraphics[align=t, height=0.18\textwidth, clip, trim = 95 43 105 32]{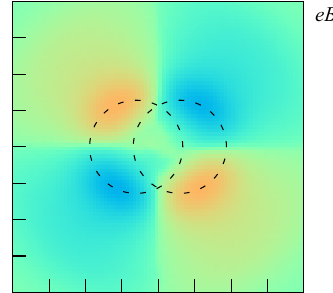}\hspace*{-5.7mm}
\includegraphics[align=t, height=0.18\textwidth, clip, trim = 95 43 105 32]{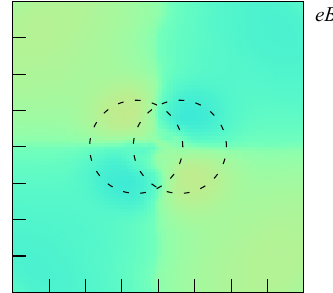}\hspace*{-5.7mm}
\includegraphics[align=t, height=0.18\textwidth, clip, trim = 95 43 105 32]{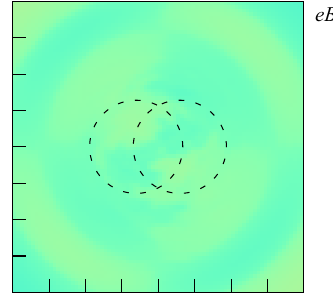}\hspace*{-11.4mm}
\includegraphics[align=t, height=0.18\textwidth, clip, trim = 95 43  55 32]{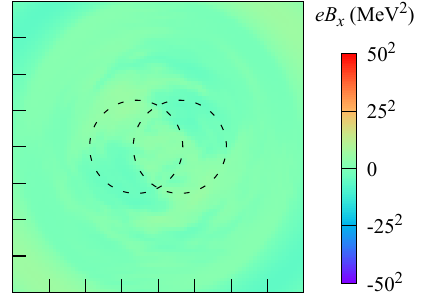}\hspace*{-6mm} \\
\vspace*{-0.7mm}
\hspace*{-33mm}
\includegraphics[align=t, height=0.18\textwidth, clip, trim = 35 43 105 32]{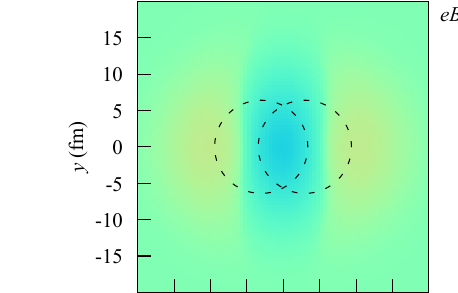}\hspace*{1.1mm}
\includegraphics[align=t, height=0.18\textwidth, clip, trim = 95 43 105 32]{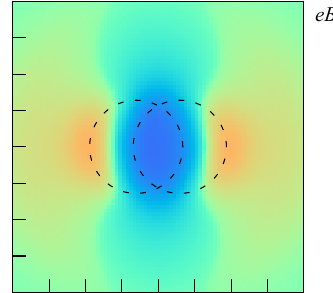}\hspace*{-5.7mm}
\includegraphics[align=t, height=0.18\textwidth, clip, trim = 95 43 105 32]{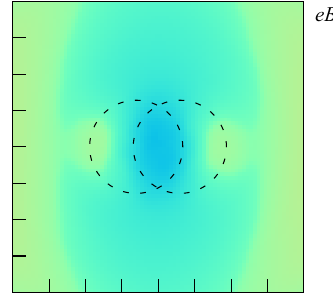}\hspace*{-5.7mm}
\includegraphics[align=t, height=0.18\textwidth, clip, trim = 95 43 105 32]{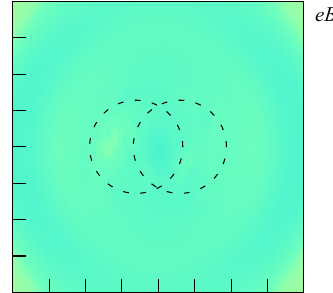}\hspace*{-11.4mm}
\includegraphics[align=t, height=0.18\textwidth, clip, trim = 95 43  55 32]{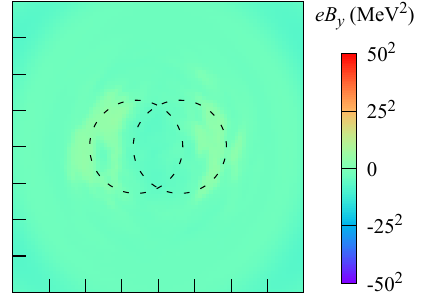}\hspace*{-6mm} \\
\vspace*{-0.7mm}
\hspace*{-33mm}
\includegraphics[align=t, height=0.2349\textwidth, clip, trim = 35 0 105 32]{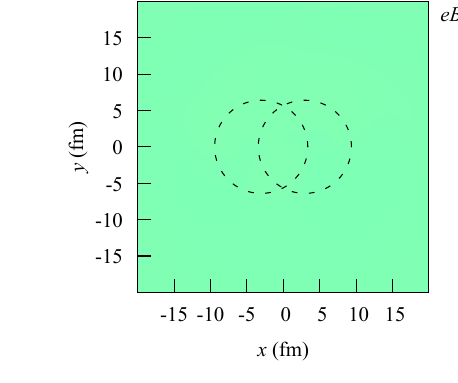}\hspace*{1.1mm}
\includegraphics[align=t, height=0.2349\textwidth, clip, trim = 95 0 105 32]{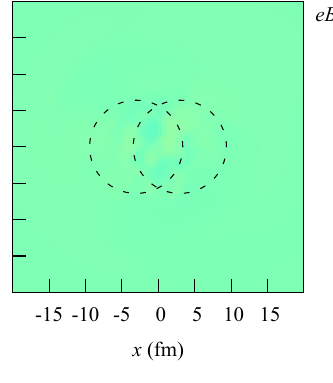}\hspace*{-5.7mm}
\includegraphics[align=t, height=0.2349\textwidth, clip, trim = 95 0 105 32]{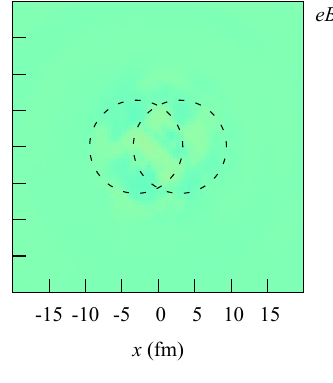}\hspace*{-5.7mm}
\includegraphics[align=t, height=0.2349\textwidth, clip, trim = 95 0 105 32]{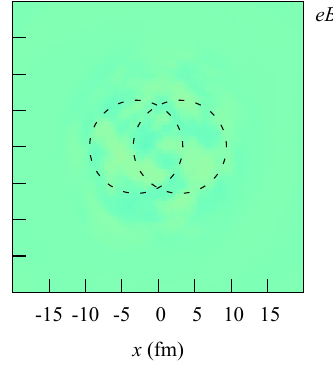}\hspace*{-11.4mm}
\includegraphics[align=t, height=0.2349\textwidth, clip, trim = 95 0  55 32]{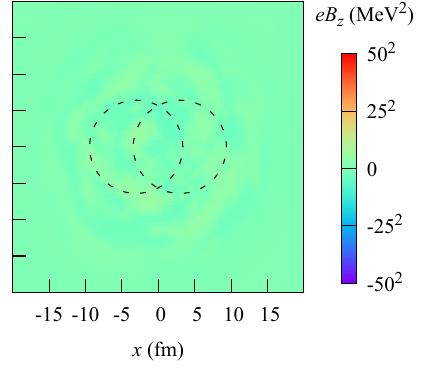}\hspace*{-6mm} \\
\caption{\label{fig:app1} The spacetime profile of the electromagnetic fields, ${\bm E}$ (top three panels) and ${\bm B}$ (bottom three panels), at a fixed collision energy $\sqsNN=4.5\;{\rm GeV}$ and an impact parameter $b=6\;{\rm fm}$, sliced at $z=0\;{\rm fm}$.  The black dashed circles indicate the locations of the colliding ions in the free-streaming limit.  }
\end{figure*}

\begin{figure*}[!t]
\flushleft{\hspace*{15.5mm}\mbox{$t=1.0\;{\rm fm}/c$ \hspace{15.6mm} $4.5\;{\rm fm}/c$ \hspace{17.6mm} $8.0\;{\rm fm}/c$ \hspace{17.6mm} $11.5\;{\rm fm}/c$ \hspace{17mm} $15.0\;{\rm fm}/c$}} \\
\vspace*{1mm}
\hspace*{-33mm}
\includegraphics[align=t, height=0.18\textwidth, clip, trim = 35 43 105 32]{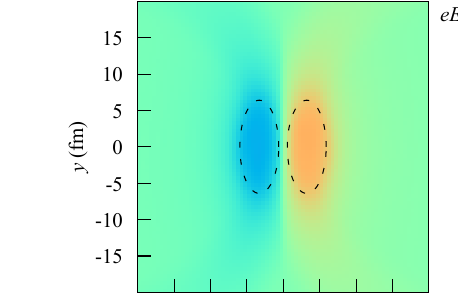}\hspace*{1.1mm}
\includegraphics[align=t, height=0.18\textwidth, clip, trim = 95 43 105 32]{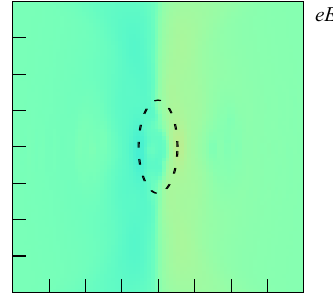}\hspace*{-5.7mm}
\includegraphics[align=t, height=0.18\textwidth, clip, trim = 95 43 105 32]{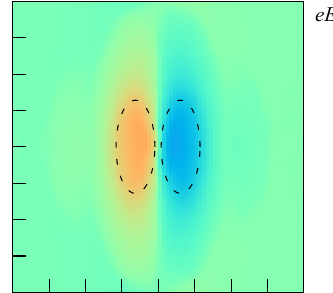}\hspace*{-5.7mm}
\includegraphics[align=t, height=0.18\textwidth, clip, trim = 95 43 105 32]{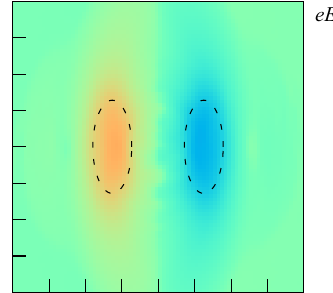}\hspace*{-11.4mm}
\includegraphics[align=t, height=0.18\textwidth, clip, trim = 95 43  55 32]{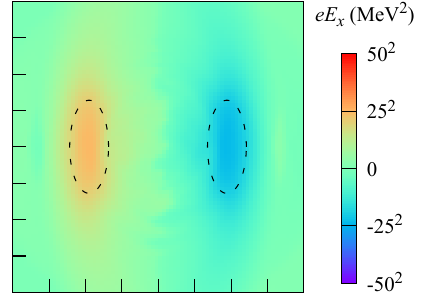}\hspace*{-6mm} \\
\vspace*{-0.7mm}
\hspace*{-33mm}
\includegraphics[align=t, height=0.18\textwidth, clip, trim = 35 43 105 32]{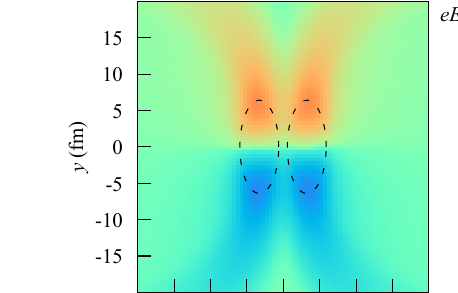}\hspace*{1.1mm}
\includegraphics[align=t, height=0.18\textwidth, clip, trim = 95 43 105 32]{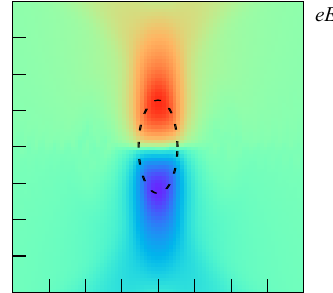}\hspace*{-5.7mm}
\includegraphics[align=t, height=0.18\textwidth, clip, trim = 95 43 105 32]{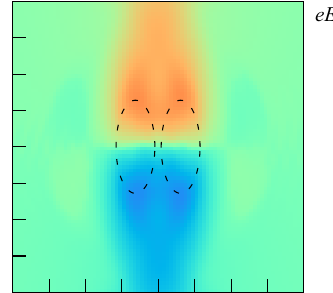}\hspace*{-5.7mm}
\includegraphics[align=t, height=0.18\textwidth, clip, trim = 95 43 105 32]{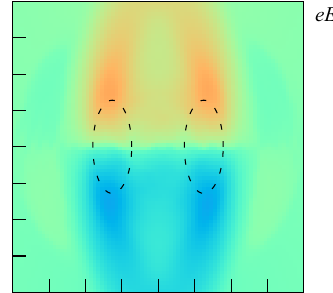}\hspace*{-11.4mm}
\includegraphics[align=t, height=0.18\textwidth, clip, trim = 95 43  55 32]{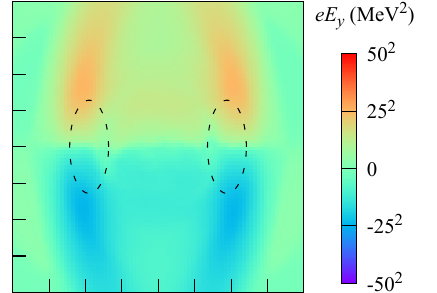}\hspace*{-6mm} \\
\vspace*{-0.7mm}
\hspace*{-33mm}
\includegraphics[align=t, height=0.18\textwidth, clip, trim = 35 43 105 32]{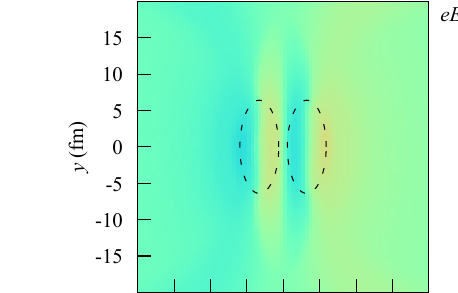}\hspace*{1.1mm}
\includegraphics[align=t, height=0.18\textwidth, clip, trim = 95 43 105 32]{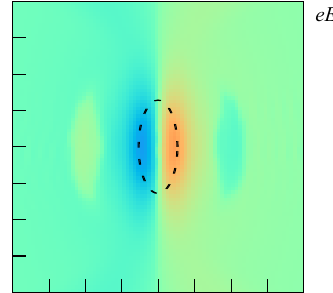}\hspace*{-5.7mm}
\includegraphics[align=t, height=0.18\textwidth, clip, trim = 95 43 105 32]{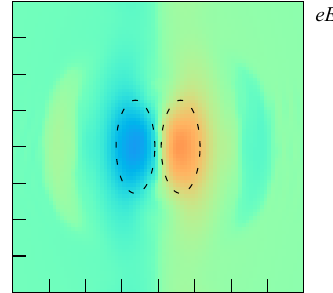}\hspace*{-5.7mm}
\includegraphics[align=t, height=0.18\textwidth, clip, trim = 95 43 105 32]{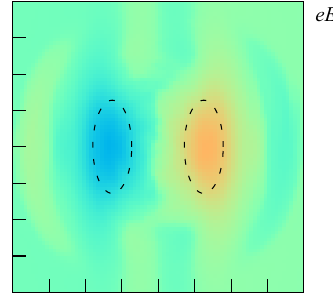}\hspace*{-11.4mm}
\includegraphics[align=t, height=0.18\textwidth, clip, trim = 95 43  55 32]{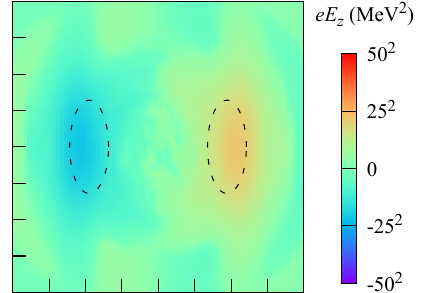}\hspace*{-6mm} \\
\vspace*{2mm}
\hspace*{-33mm}
\includegraphics[align=t, height=0.18\textwidth, clip, trim = 35 43 105 32]{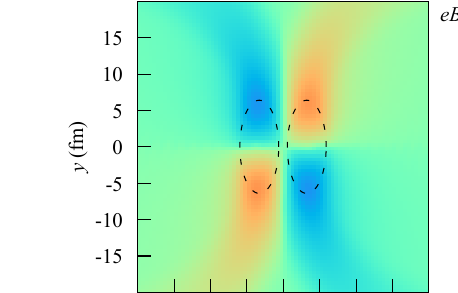}\hspace*{1.1mm}
\includegraphics[align=t, height=0.18\textwidth, clip, trim = 95 43 105 32]{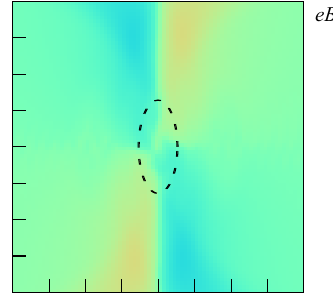}\hspace*{-5.7mm}
\includegraphics[align=t, height=0.18\textwidth, clip, trim = 95 43 105 32]{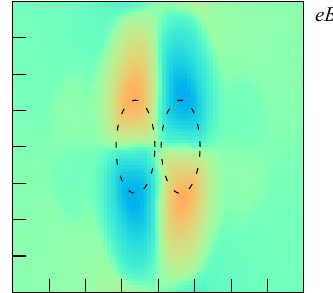}\hspace*{-5.7mm}
\includegraphics[align=t, height=0.18\textwidth, clip, trim = 95 43 105 32]{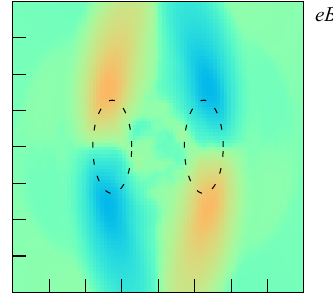}\hspace*{-11.4mm}
\includegraphics[align=t, height=0.18\textwidth, clip, trim = 95 43  55 32]{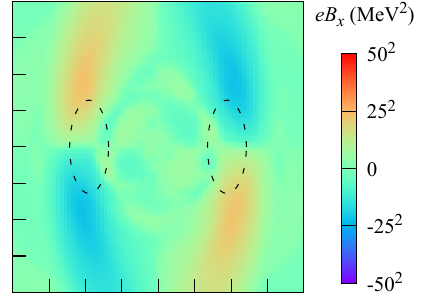}\hspace*{-6mm} \\
\vspace*{-0.7mm}
\hspace*{-33mm}
\includegraphics[align=t, height=0.18\textwidth, clip, trim = 35 43 105 32]{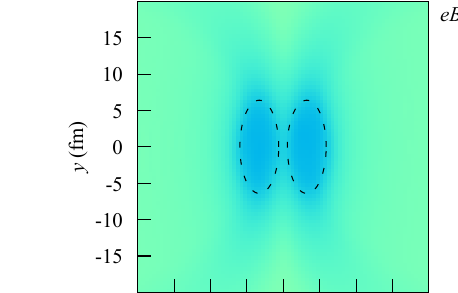}\hspace*{1.1mm}
\includegraphics[align=t, height=0.18\textwidth, clip, trim = 95 43 105 32]{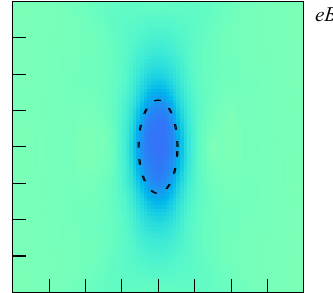}\hspace*{-5.7mm}
\includegraphics[align=t, height=0.18\textwidth, clip, trim = 95 43 105 32]{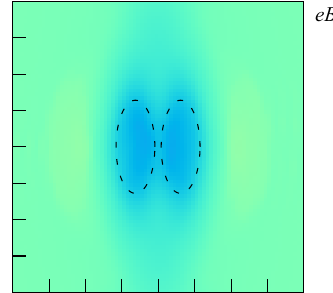}\hspace*{-5.7mm}
\includegraphics[align=t, height=0.18\textwidth, clip, trim = 95 43 105 32]{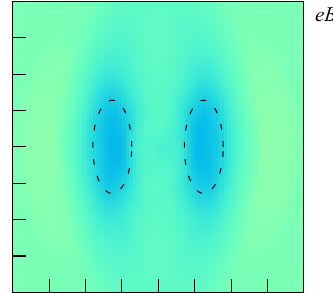}\hspace*{-11.4mm}
\includegraphics[align=t, height=0.18\textwidth, clip, trim = 95 43  55 32]{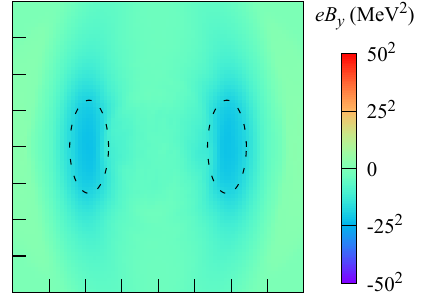}\hspace*{-6mm} \\
\vspace*{-0.7mm}
\hspace*{-33mm}
\includegraphics[align=t, height=0.2349\textwidth, clip, trim = 35 0 105 32]{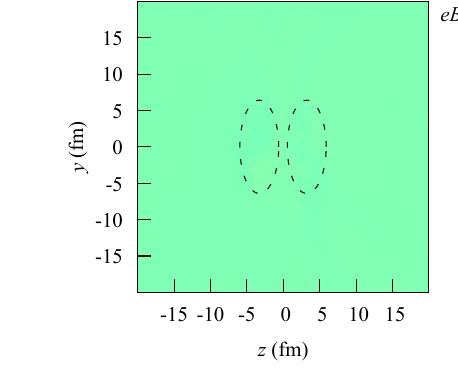}\hspace*{1.1mm}
\includegraphics[align=t, height=0.2349\textwidth, clip, trim = 95 0 105 32]{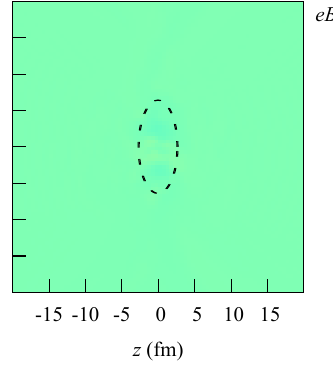}\hspace*{-5.7mm}
\includegraphics[align=t, height=0.2349\textwidth, clip, trim = 95 0 105 32]{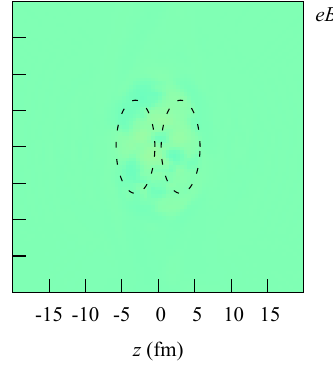}\hspace*{-5.7mm}
\includegraphics[align=t, height=0.2349\textwidth, clip, trim = 95 0 105 32]{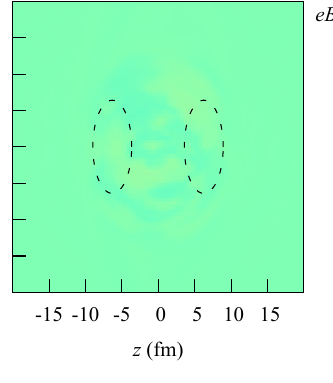}\hspace*{-11.4mm}
\includegraphics[align=t, height=0.2349\textwidth, clip, trim = 95 0  55 32]{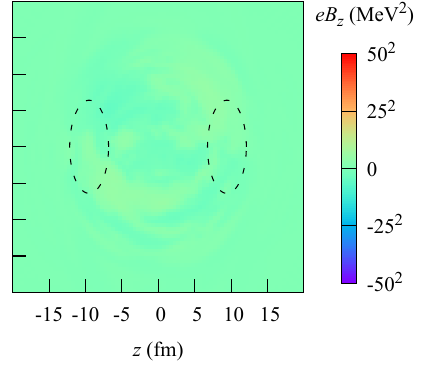}\hspace*{-6mm} \\
\caption{\label{fig:app2} A same plot as Fig.~\ref{fig:app1} but sliced at $x=0\;{\rm fm}$.  }
\end{figure*}

\begin{figure*}[!t]
\flushleft{\hspace*{15.5mm}\mbox{$t=1.0\;{\rm fm}/c$ \hspace{15.6mm} $4.5\;{\rm fm}/c$ \hspace{17.6mm} $8.0\;{\rm fm}/c$ \hspace{17.6mm} $11.5\;{\rm fm}/c$ \hspace{17mm} $15.0\;{\rm fm}/c$}} \\
\vspace*{1mm}
\hspace*{-33mm}
\includegraphics[align=t, height=0.18\textwidth, clip, trim = 35 43 105 32]{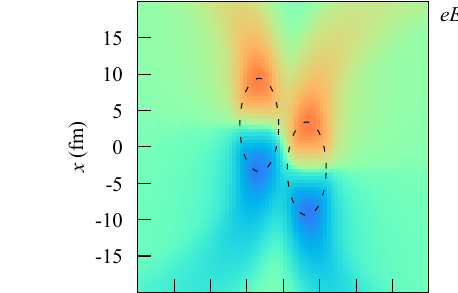}\hspace*{1.1mm}
\includegraphics[align=t, height=0.18\textwidth, clip, trim = 95 43 105 32]{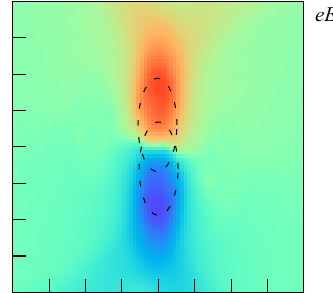}\hspace*{-5.7mm}
\includegraphics[align=t, height=0.18\textwidth, clip, trim = 95 43 105 32]{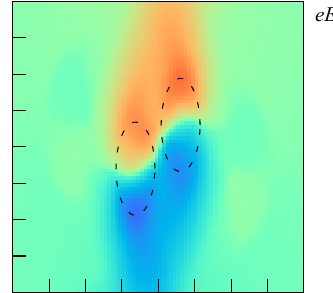}\hspace*{-5.7mm}
\includegraphics[align=t, height=0.18\textwidth, clip, trim = 95 43 105 32]{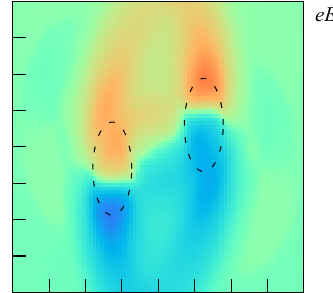}\hspace*{-11.4mm}
\includegraphics[align=t, height=0.18\textwidth, clip, trim = 95 43  55 32]{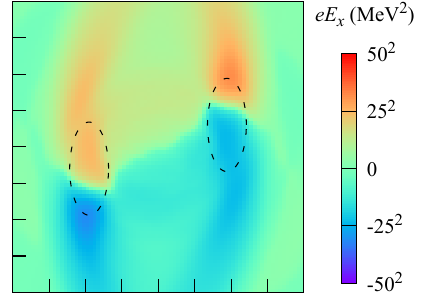}\hspace*{-6mm} \\
\vspace*{-0.7mm}
\hspace*{-33mm}
\includegraphics[align=t, height=0.18\textwidth, clip, trim = 35 43 105 32]{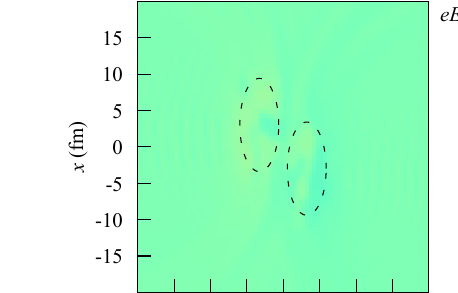}\hspace*{1.1mm}
\includegraphics[align=t, height=0.18\textwidth, clip, trim = 95 43 105 32]{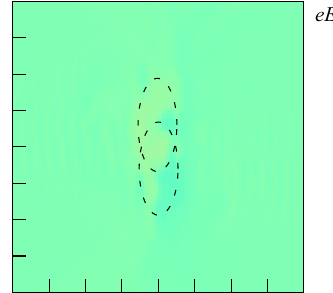}\hspace*{-5.7mm}
\includegraphics[align=t, height=0.18\textwidth, clip, trim = 95 43 105 32]{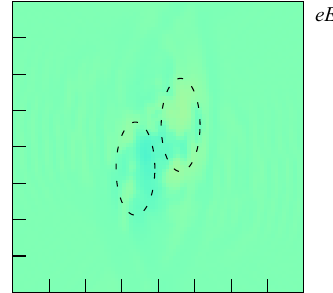}\hspace*{-5.7mm}
\includegraphics[align=t, height=0.18\textwidth, clip, trim = 95 43 105 32]{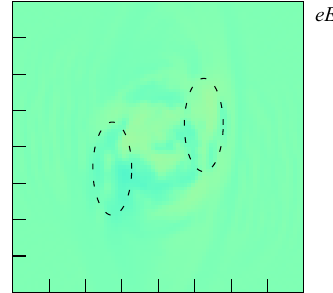}\hspace*{-11.4mm}
\includegraphics[align=t, height=0.18\textwidth, clip, trim = 95 43  55 32]{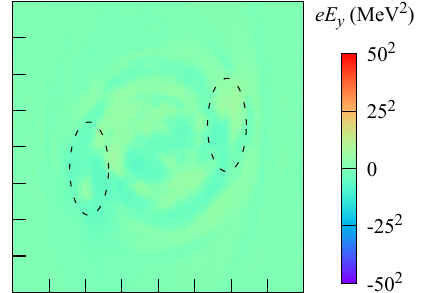}\hspace*{-6mm} \\
\vspace*{-0.7mm}
\hspace*{-33mm}
\includegraphics[align=t, height=0.18\textwidth, clip, trim = 35 43 105 32]{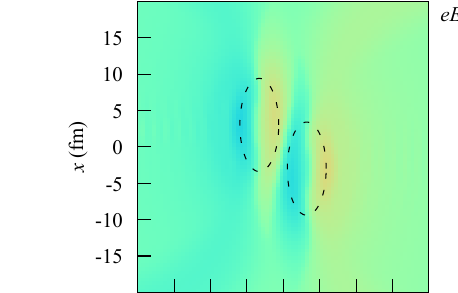}\hspace*{1.1mm}
\includegraphics[align=t, height=0.18\textwidth, clip, trim = 95 43 105 32]{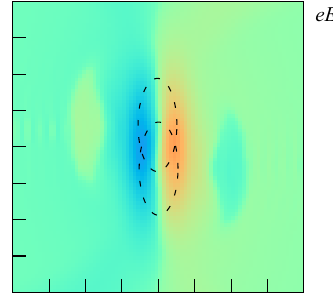}\hspace*{-5.7mm}
\includegraphics[align=t, height=0.18\textwidth, clip, trim = 95 43 105 32]{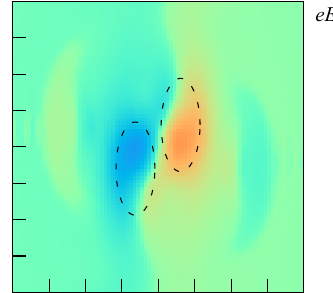}\hspace*{-5.7mm}
\includegraphics[align=t, height=0.18\textwidth, clip, trim = 95 43 105 32]{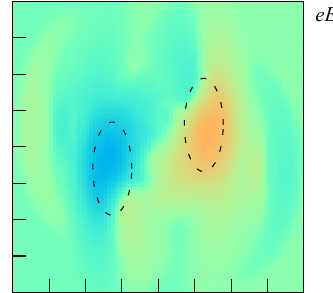}\hspace*{-11.4mm}
\includegraphics[align=t, height=0.18\textwidth, clip, trim = 95 43  55 32]{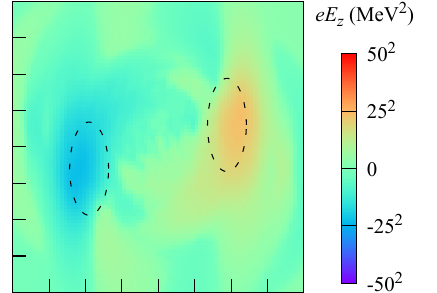}\hspace*{-6mm} \\
\vspace*{2mm}
\hspace*{-33mm}
\includegraphics[align=t, height=0.18\textwidth, clip, trim = 35 43 105 32]{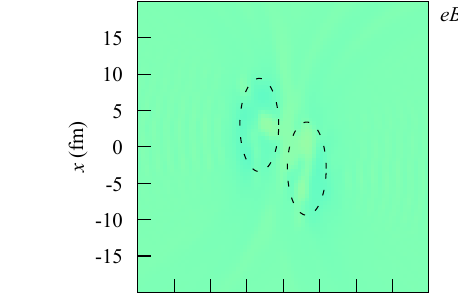}\hspace*{1.1mm}
\includegraphics[align=t, height=0.18\textwidth, clip, trim = 95 43 105 32]{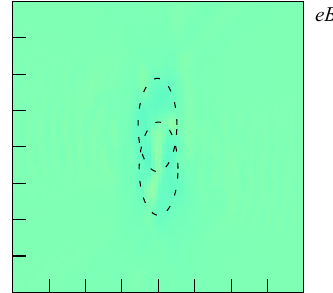}\hspace*{-5.7mm}
\includegraphics[align=t, height=0.18\textwidth, clip, trim = 95 43 105 32]{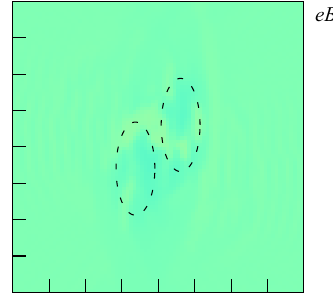}\hspace*{-5.7mm}
\includegraphics[align=t, height=0.18\textwidth, clip, trim = 95 43 105 32]{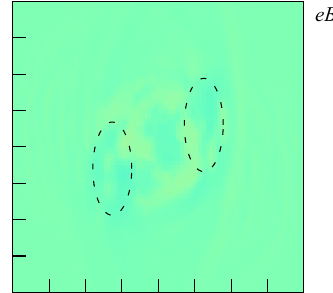}\hspace*{-11.4mm}
\includegraphics[align=t, height=0.18\textwidth, clip, trim = 95 43  55 32]{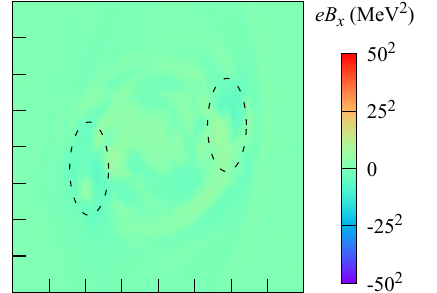}\hspace*{-6mm} \\
\vspace*{-0.7mm}
\hspace*{-33mm}
\includegraphics[align=t, height=0.18\textwidth, clip, trim = 35 43 105 32]{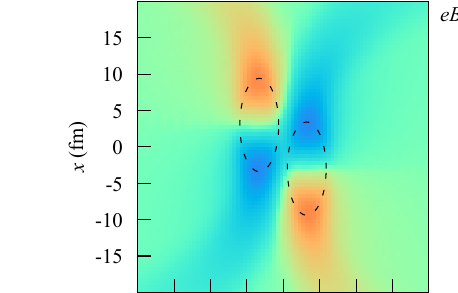}\hspace*{1.1mm}
\includegraphics[align=t, height=0.18\textwidth, clip, trim = 95 43 105 32]{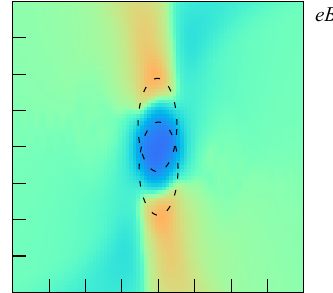}\hspace*{-5.7mm}
\includegraphics[align=t, height=0.18\textwidth, clip, trim = 95 43 105 32]{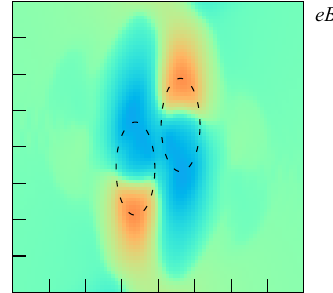}\hspace*{-5.7mm}
\includegraphics[align=t, height=0.18\textwidth, clip, trim = 95 43 105 32]{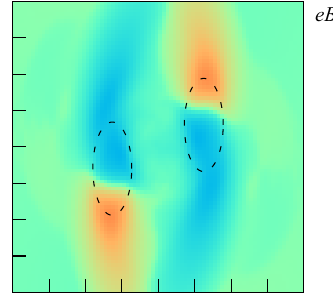}\hspace*{-11.4mm}
\includegraphics[align=t, height=0.18\textwidth, clip, trim = 95 43  55 32]{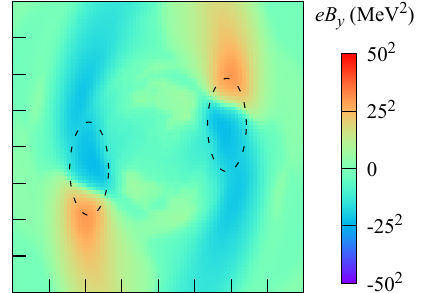}\hspace*{-6mm} \\
\vspace*{-0.7mm}
\hspace*{-33mm}
\includegraphics[align=t, height=0.2349\textwidth, clip, trim = 35 0 105 32]{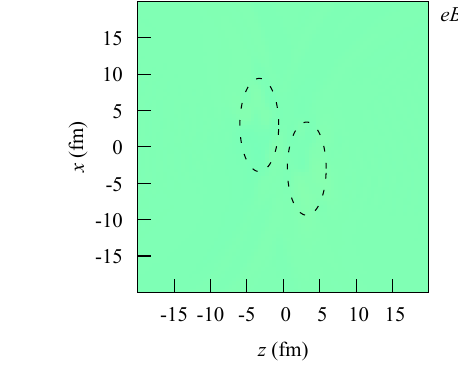}\hspace*{1.1mm}
\includegraphics[align=t, height=0.2349\textwidth, clip, trim = 95 0 105 32]{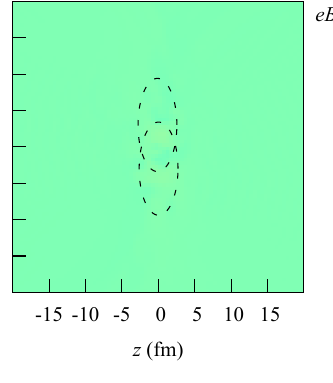}\hspace*{-5.7mm}
\includegraphics[align=t, height=0.2349\textwidth, clip, trim = 95 0 105 32]{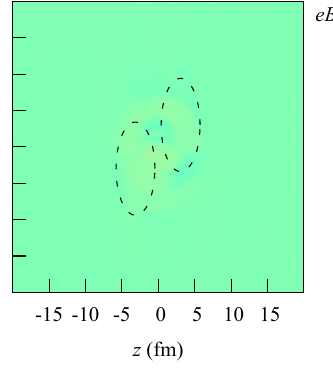}\hspace*{-5.7mm}
\includegraphics[align=t, height=0.2349\textwidth, clip, trim = 95 0 105 32]{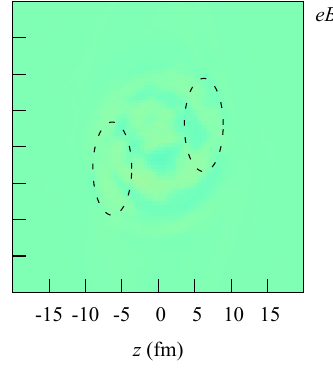}\hspace*{-11.4mm}
\includegraphics[align=t, height=0.2349\textwidth, clip, trim = 95 0  55 32]{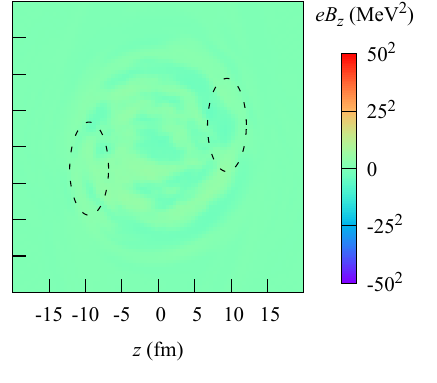}\hspace*{-6mm} \\
\caption{\label{fig:app3} A same plot as Figs.~\ref{fig:app1} and \ref{fig:app2} but sliced at $y=0\;{\rm fm}$.  }
\end{figure*}

In the main text, I focused on the spacetime evolution of the electromagnetic Lorentz invariants, ${\mathcal F}$ and ${\mathcal G}$, rather than each of the electric and magnetic components, ${\bm E}$ and ${\bm B}$, in order to make the figures and discussions simpler.  In Figs.~\ref{fig:app1}\,-\,\ref{fig:app3}, for interested readers, I display the correspondence of Fig.~\ref{fig:1} for ${\bm E}$ and ${\bm B}$.

\bibliography{bib}
\end{document}